\documentclass[twocolumn]{aastex7}

\begin{document}

\title[Breathing mode in low-mass galaxies]{Testing the Stellar Feedback-driven Breathing Mode in Low-mass Galaxies with Gas Kinematics}

\author[0000-0001-7729-6629]{Yifei Luo}
\affiliation{Lawrence Berkeley National Laboratory, 1 Cyclotron Road, Berkeley, CA 94720, USA}
\affiliation{Department of Astronomy and Astrophysics, University of California, Santa Cruz, 1156 High Street, Santa Cruz, CA 95064 USA}
\email[show]{yifeiluo@lbl.gov}

\author[0000-0001-9833-6183]{Joseph Wick}
\affil{Department of Astronomy and Astrophysics, University of California, Santa Cruz, 1156 High Street, Santa Cruz, CA 95064 USA}
\email{JosephWick@outlook.com}

\author[0000-0002-3677-3617]{Alexie Leauthaud}
\affil{Department of Astronomy and Astrophysics, University of California, Santa Cruz, 1156 High Street, Santa Cruz, CA 95064 USA}
\email{alexie@ucsc.edu}

\author[0000-0003-0603-8942]{Andrew Wetzel}
\affil{Department of Physics \& Astronomy, University of California, Davis, CA 95616, USA}
\email{awetzel@ucdavis.edu}

\author[0000-0001-5860-3419]{Tucker Jones}
\affil{Department of Physics \& Astronomy, University of California, Davis, CA 95616, USA}
\email{tdjones@ucdavis.edu}

\author[0000-0002-0332-177X]{Erin Kado-Fong}
\affil{Physics Department, Yale Center for Astronomy \& Astrophysics, PO Box 208120, New Haven, CT 06520, USA}
\email{erin.kado-fong@yale.edu}

\author[0000-0003-1385-7591]{Song Huang}
\affil{Department of Astronomy, Tsinghua University, Beijing 100084, China}
\email{shuang@tsinghua.edu.cn}

\author[0009-0005-0700-5271]{Xinjun Chen}
\affil{Department of Astronomy and Astrophysics, University of California, Santa Cruz, 1156 High Street, Santa Cruz, CA 95064 USA}
\affil{Department of Physics and Astronomy, University of Missouri, Columbia, MO 65211, USA}
\email{xcgc6@missouri.edu}

\author[0000-0002-2897-6326]{Conghao Zhou}
\affil{Department of Physics, University of California, Santa Cruz, 1156 High Street, Santa Cruz, CA 95064 USA}
\email{zhou.conghao@ucsc.edu}

\author[0000-0001-9592-4190]{Jiaxuan Li}
\affil{Department of Astrophysical Sciences, Princeton University, 4 Ivy Lane, Princeton, NJ 08544, USA}
\email{jiaxuanl@princeton.edu}

\begin{abstract}

Hydrodynamic simulations have proposed that stellar feedback and bursty star-formation can produce dark matter cores in low-mass galaxies. A key prediction is that feedback-driven gas outflow and inflow cycles can lead to ``breathing modes'' (rapid fluctuations in the global gravitational potential) which drive correlated variations in galaxy size, kinematics, and star-formation rate. In this paper, we test the dynamical effects of feedback-driven breathing modes using a sample of 103 star-forming low-mass galaxies with stellar masses between $7.9<\rm \log M_*/M_\odot<9.6$ and $0.02<z<0.19$. We measure ionized gas velocity dispersions from H$\alpha$ emission lines and compare them to mock observations from the FIRE-2 simulations. We compare gas velocity dispersions ($\rm \sigma_{gas}$), stellar masses, and specific star-formation rates (sSFR). We find a positive correlation between gas velocity dispersion residuals at fixed stellar masses ($\rm \Delta\sigma_{gas}$) and sSFR in both data and simulations. However, the relation is tighter in FIRE-2 compared to the data. FIRE-2 produces more low-sSFR galaxies compared to our observational sample, however, the sSFR distributions agree after limiting both samples to a minimum sSFR. A deeper and more complete photometric sample further indicates that observed low-mass galaxies could span the full range of sSFR predicted in the FIRE-2 simulations. Our results support the existence of short-timescale dynamical effects driven by gas outflow and inflow cycles in low-mass galaxies and motivate additional tests of the breathing mode. 

\end{abstract}


\section{Introduction}
Low-mass (``dwarf'') galaxies have been studied for decades as probes of the nature of dark matter and the interplay between dark matter and baryonic physics. Observations of low-mass ($\rm M_*<10^{9.5}M_\odot$) galaxies indicate that their dark-matter halos exhibit a diversity of density profiles, with many being consistent with nearly constant density ``cores'' \citep{McGaugh:2001,deBlok:2008}, while cold dark matter (CDM) only models uniformly predict steep central profiles \citep[``cusps''][]{Navarro:1996,Navarro:2010}. This core-cusp/diversity problem is one of the most significant and long-standing challenges to CDM \citep[e.g.][]{Bullock:2017}. Such problems could be explained by alternative dark matter models such as self-interacting dark matter ``SIDM", or by more detailed baryonic processes within $\Lambda$CDM \citep[e.g.][]{Vogelsberger:2014,Fry:2015,Kamada:2017,Robles:2017,Fitts:2019,Nadler:2021,Sales:2022}. 

Recent work has shown that stellar feedback could be capable of altering the dark-matter distribution inside dwarf galaxies, resolving the core-cusp/diversity problem \citep[e.g.][]{Governato:2010, Pontzen:2012,Chan:2015,El-Badry:2016, Hopkins:2020}. In this scenario, stellar feedback in gas-rich dwarf galaxies can drive gas outflows, shallowing the gravitational potential. Over time, these non-adiabatic fluctuations in the potential can heat the orbits of dark matter (and stars). The gas then can cool and fall back into the galaxy, repeating the cycle. Over multiple starbursts, this process systematically lowers the density of dark matter within the galaxy. Multiple studies have shown that this process acts most significantly at M$_{\rm *} \approx 10^{7-9} \rm M_{\odot}$ \citep[e.g.][]{Mashchenko:2008,Governato:2012,Pontzen:2012,Teyssier:2013,Lazar:2020,Tollet:2016}.

However, it remains unclear whether the way that these simulations model stellar feedback is realistic, and in particular, if stellar feedback is strong enough to induce significant changes to the dark matter distribution. Therefore, observations are needed to constrain these effects.

While promising, a challenge has been identifying unique, direct observational tests of feedback-driven core formation. A recent series of papers \citep[e.g.][]{Christensen:2016,El-Badry:2016,El-Badry:2017, Read:2019} used cosmological zoom-in simulations to predict a new observational test: the ``breathing mode" in low-mass galaxies. The prediction is that baryons respond to feedback-driven potential fluctuations similarly to dark matter, leading to observable correlations between the recent specific star-formation rate (sSFR), stellar and gas kinematics, galaxy sizes, and dark-matter coring. Specifically, gas cooling into the galaxy center first leads to a deep potential, compact stellar size, high sSFR, and high velocity dispersion \citep{El-Badry:2016,El-Badry:2017}. A feedback-driven outflow then produces a shallow potential well, large stellar size, low sSFR, and low $\rm \sigma_{gas}$. Each of these cycles occurs over a timescale of a few hundred million years within a single galaxy, persisting down to $z\sim0$. The bursty star-formation has also been proposed to be a possible formation mechanism of ultra diffuse galaxies \citep[e.g.][]{Chan:2018, DiCintio:2017}.

A number of observational and theoretical studies have tested the correlations between star formation activity, galaxy size, and kinematics, driven by stellar feedback. \citet{Patel:2018} and \citet{Emami:2021} tested the size fluctuations caused by stellar feedback with observations and simulations, and found the sSFR-size relation is sensitive to the timescales of the bursty star-formation. \citet{Cicone:2016} and \citet{Pelliccia:2020} used large samples drawn from the Sloan Digital Sky Survey \citep[SDSS, $\rm \sigma_{inst}\sim70\ km/s$,][]{Abazajian:2009} and Large Early Galaxy Astrophysics Census \citep[LEGA-C, $\rm \sigma_{inst}\sim50\ km/s$,][]{vanderWel:2016} to test the sSFR-$\rm \sigma$ relation in galaxies with $\rm \log M_*/M_\odot<10$. \citet{Wang:2024} found a negative correlation between the galaxy stellar population age $\rm D_{n}4000$ (which is sensitive to star-formation) and galaxy rotational support indicator $\rm \lambda_{Re}$ for low-mass galaxies with $9.5<\rm \log M_*/M_\odot<10.0$, i.e. galaxies in this mass range with younger stellar population (or higher specific star-formation rates) tend to be more velocity dispersion supported systems. Previous studies also reported positive relations between $\rm \sigma_{gas}$ and SFR or $\rm \Sigma_{SFR}$ in galaxies at $\rm \log M_*/M_\odot>8.5$ \citep{Yu:2019,Varidel:2020,Law:2022}, using Integral-Field Unit (IFU) data from MaNGA \citep[$\rm \sigma_{inst}\sim 50-80\ km/s$,][]{Bundy:2015} and SAMI \citep[$\rm \sigma_{inst}\sim 30\ km/s$,][]{Croom:2012}. However, those surveys are not optimal for testing the sSFR-$\rm \sigma$ relation in low-mass galaxies, not only due to survey depth, but also due to spectral resolution, since the typical gas velocity dispersion for a $\rm \log M_*/M_\odot=8$ galaxy is 15-30 km/s \citep[e.g.][]{Kassin:2012,Cortese:2014}. Therefore, to test the ``breathing mode" stellar feedback in low-mass galaxies with kinematics, a sample of galaxies with high enough spectral resolution is needed, in order to accurately measure the kinematics in those low-mass systems. 

\citet{Hirtenstein:2019} tested the relation between sSFR and $\rm \sigma_{gas}$ with 17 star-forming galaxies with stellar masses $8<\rm \log M_*/M_\odot<9.8$ at redshifts $1.2 < z < 2.3$, and directly compared the results with the FIRE-2 simulations \citep{Hopkins:2018}. They took advantage of massive galaxy clusters as gravitational lenses, and utilized adaptive optics on the OH-Suppressing Infrared Imaging Spectrograph (OSIRIS) at W. M. Keck Observatory, in order to measure the high quality kinematic data ($\rm \sigma_{inst}\sim35 km/s$) for low-mass galaxies at $z\sim2$. Their results support ``breathing mode" stellar feedback with good agreement between observations and FIRE-2 simulations. Nevertheless, their sample size is limited and subject to selection effects in both surface brightness and SFR. In fact, the average SFR in galaxies at $z\sim2$ is systematically higher than $z\sim0$ \citep{Madau:2014}. Therefore, a larger sample of low-mass galaxies spanning a wider range of SFR at low redshifts, with precise gas kinematics measurements, is still needed in order to directly compare with the predictions made by simulations at $z\sim0$ \citep[e.g.][]{El-Badry:2016}.

In this paper, we build a sample of 103 low-mass galaxies with $7.9<\rm \log M_*/M_\odot<9.6$ at $0.02<z<0.19$ with Keck/DEIMOS spectroscopic observations and 30-band photometry from UV to IR from the COSMOS2015 catalog \citep{Laigle:2016}, to test the "breathing mode" with observed dwarf galaxies. This sample has not only larger sample size, but also higher completeness in mass and SFR compared to the $z\sim 2$ low-mass galaxies sample of \citet{Hirtenstein:2019}. In addition, it is at lower redshifts, which is closer to the dwarf galaxies in the nearby universe that the original ``core-cusp'' problem was found. It is essential to test whether the ``breathing mode" stellar feedback could also be observed in the low-redshift universe.

This paper is organized as follows. The data and sample used to test the "breathing mode" from observations and simulations are described in Section~\ref{data}. In Section~\ref{results} we show the gas kinematics and star-formation activities from observations and compare them with simulations. In Section~\ref{discussion} we discuss our results. Section~\ref{conclusions} presents a summary and our conclusions. We adopt a flat $\Lambda$CDM cosmology with $H_0$ = 70 km $\rm s^{-1}\ Mpc^{-1}$, $\Omega_{\rm m}$ = 0.3 and $\Omega_\Lambda$ = 0.7.

\section{Data and Method}\label{data}

In this section, we describe both the observational and simulation data, and describe how we analyze the data to make direct comparison between observations and simulations.

\begin{figure}
\begin{center}
\includegraphics[width=8cm]{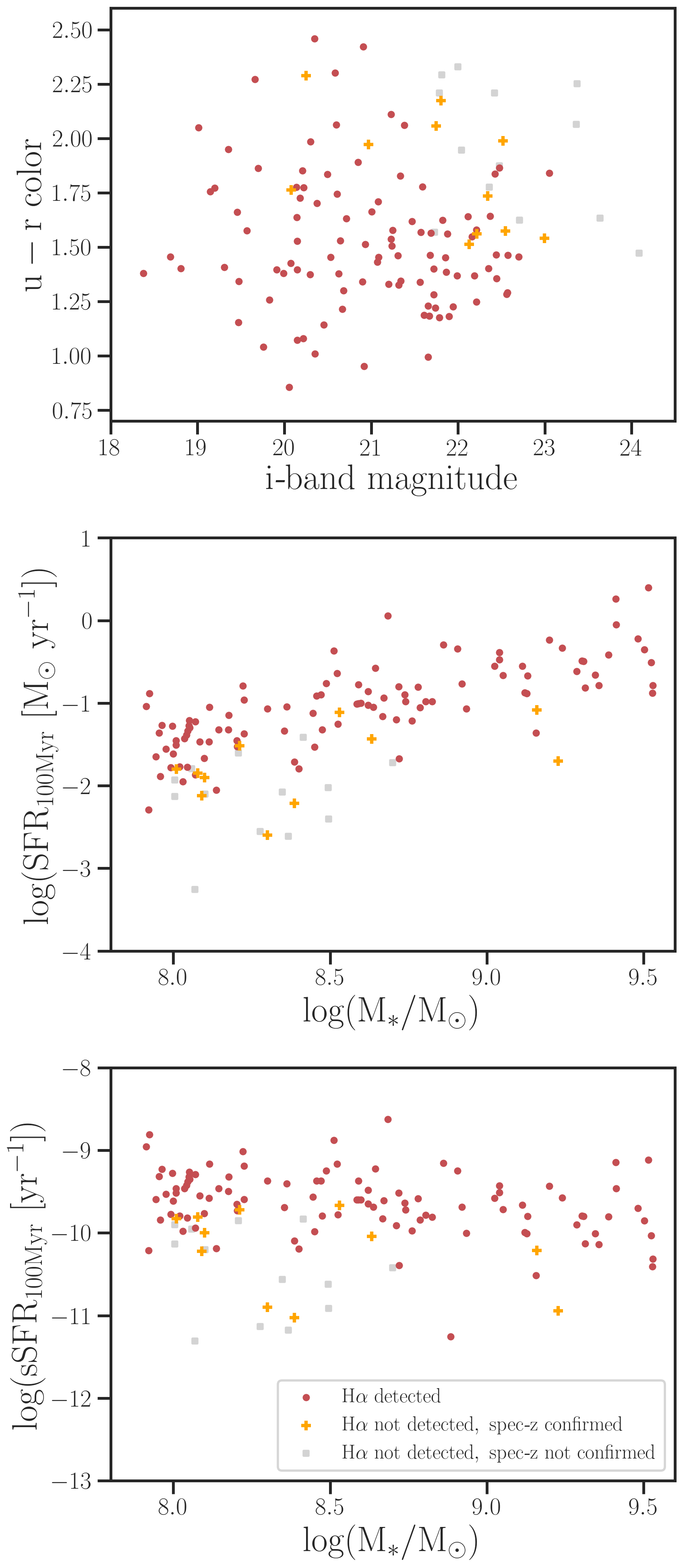}
\caption{Top: $u-r$ color vs. $i$-band magnitude for all the observed low-mass galaxies with Keck/DEIMOS ($7.9<\rm \log M_*/M_\odot<9.6$ at $0.02<z<0.19$, selected from COSMOS2015 catalog). Red points show galaxies that have H$\alpha$ emission lines detected from our Keck/DEIMOS observations. Orange crosses show galaxies confirmed with other spectroscopic observations but without H$\alpha$ measurements from our Keck/DEIMOS observations. Gray squares are galaxies that don't have any successful spectroscopic confirmation. Middle: star-formation rates vs. stellar masses. Bottom: specific star-formation rates vs. stellar masses. Our H$\alpha$ detected sample is highly complete for star-forming galaxies with $\rm log(sSFR)>-10.5\ yr^{-1}$.}
\label{ha_detection}
\end{center}
\end{figure}

\begin{figure}
\begin{center}
\includegraphics[width=8cm]{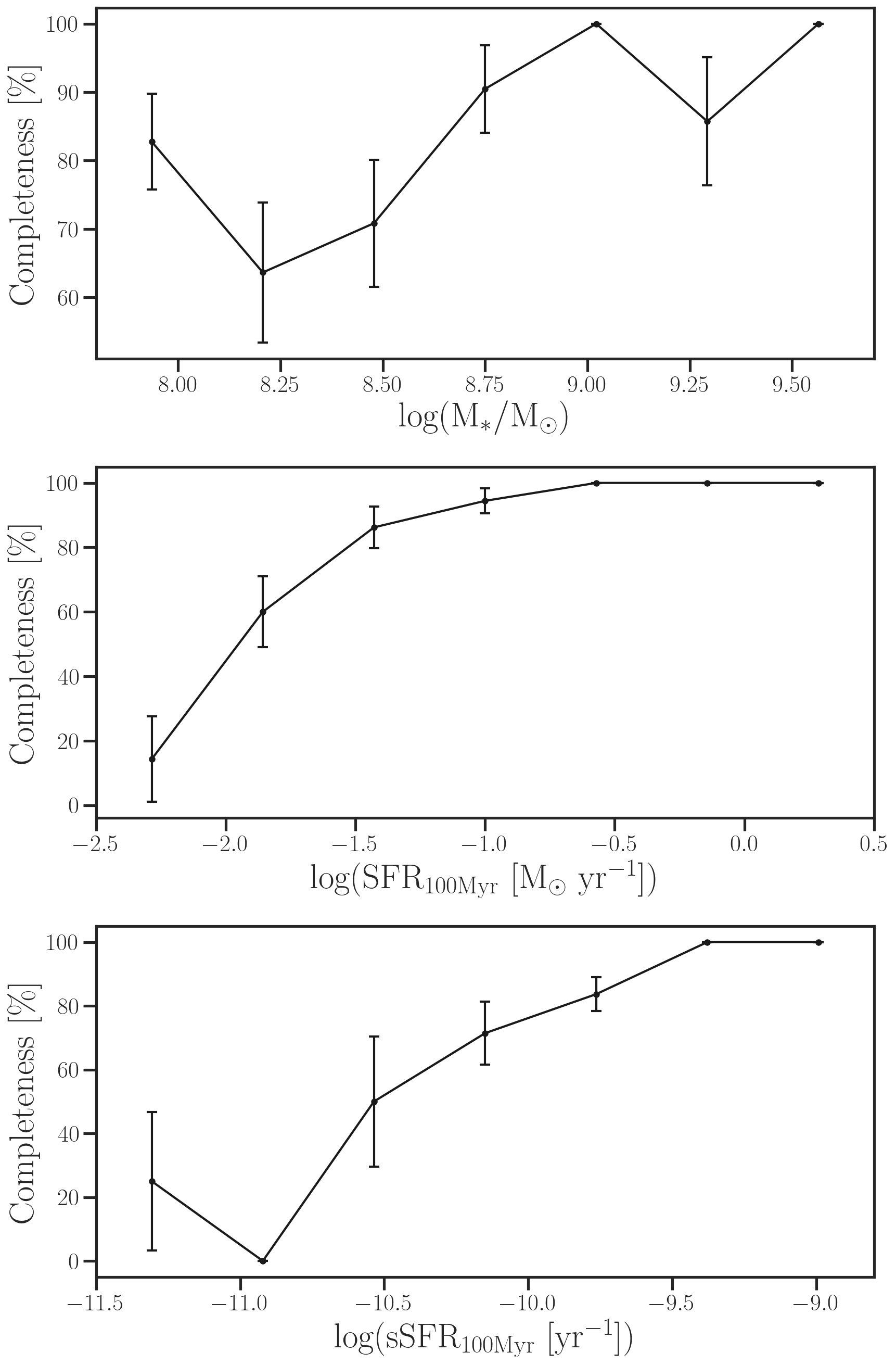}
\caption{Completeness as a function of stellar masses (top), SFR (middle), and sSFR (bottom) for our H$\alpha$ detected sample. Error bars represent binomial uncertainties based on the number of galaxies in each bin. Overall, our H$\alpha$ detected sample is 81.6\% complete compared to the much deeper photometric sample. The completeness generally increases with stellar masses, star-formation rates, and specific star-formation rates. }
\label{completeness}
\end{center}
\end{figure}

\begin{figure*}
\begin{center}
\includegraphics[width=16cm]{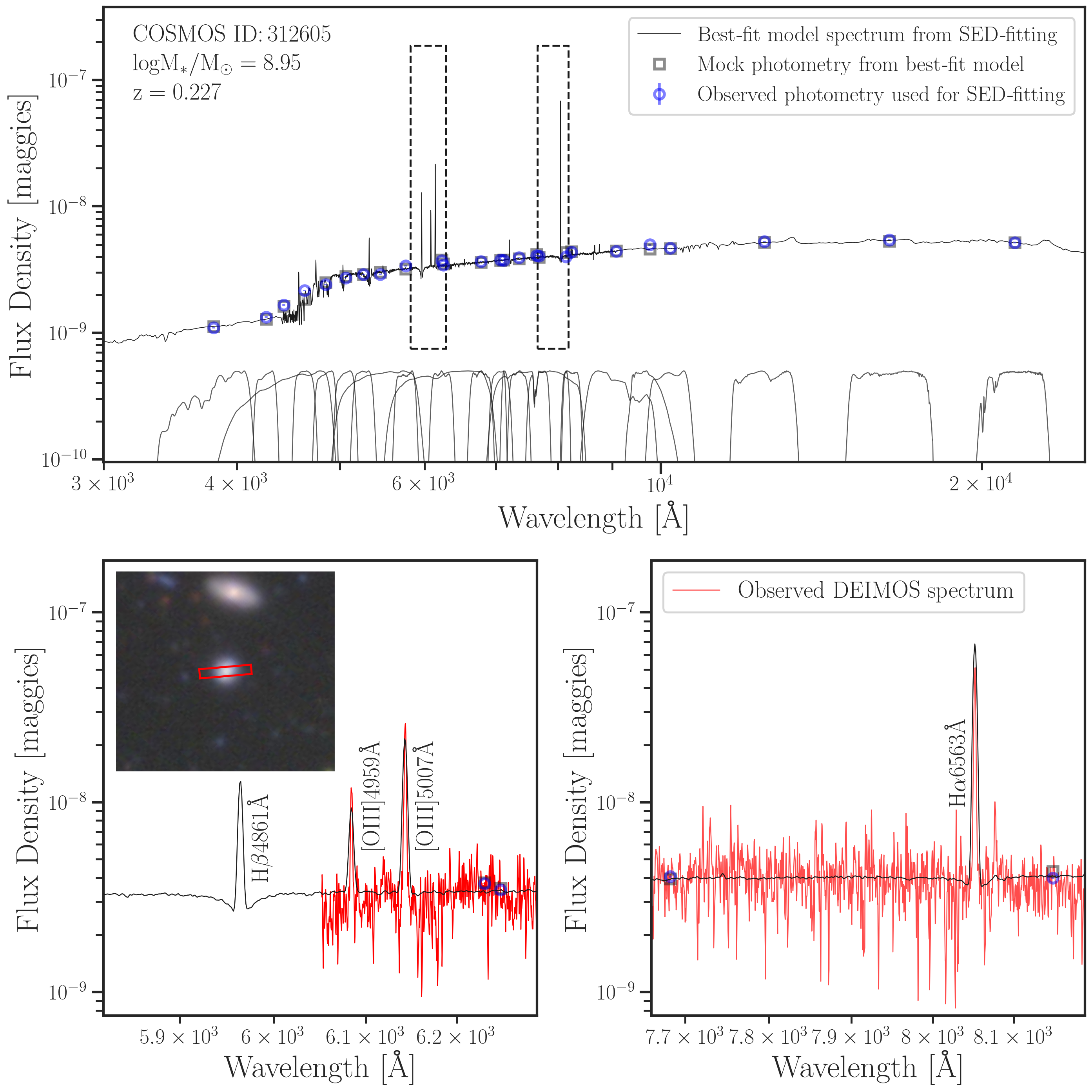}
\caption{Top: spectral energy distribution of an example dwarf galaxy (COSMOS2015 ID: 312605) in our sample. Blue circles are the multi-band photometry obtained from the COSMOS2015 catalog and used for SED fitting with \texttt{Prospector}. The black spectrum and gray boxes show the best-fit model spectrum and mock photometry from the SED fitting. Gray curves on the bottom show the filter transmission curves for each flux measurement. The lower two panels show the zoom-in regions around [OIII] (left panel) and H$\alpha$ (right panel). The red curves are the observed Keck/DEIMOS spectrum for this object. The lower left panel includes the HSC color image of this galaxy. The red rectangle on the galaxy image indicates the slit used for the Keck/DEIMOS observation. These data allow us to derive key properties relevant for this study, including stellar masses, star-formation rates, velocity dispersions.}
\label{example_galaxy}
\end{center}
\end{figure*}

\subsection{Observational data}

To test the feedback-driven breathing mode with observations, we need a sample of dwarf galaxies with both deep imaging data in order to detect the dwarf galaxies, and high resolution spectroscopic data to measure the gas kinematics. We build this sample by selecting dwarf galaxies with $7.9<\rm \log M_*/M_\odot<9.6$ from the COSMOS2015 catalog, and then obtain high spectral resolution spectroscopic data with Keck/DEIMOS. 

\subsubsection{COSMOS2015 catalog}

We first select a sample of dwarf galaxies from the COSMOS2015 catalog. The COSMOS2015 catalog \citep{Laigle:2016} provides deep and multi-wavelength photometric data covering 2 square degrees in the COSMOS field \citep{Scoville:2007}. It combines photometric data in NUV \citep[GALEX,][]{Zamojski:2007}, optical \citep[CFHT/MegaCam, Subaru COSMOS-20, Hyper-Suprime-Cam][]{Taniguchi:2007,Taniguchi:2015,Miyazaki:2018}, near-IR \citep[UltraVISTA][]{McCracken:2012}, mid-IR \citep[MIPS/Spitzer][]{LeFloch:2009}, and far-IR \citep[PACS/Herschel and SPIRE/Herschel][]{Lutz:2011,Oliver:2012}. The COSMOS2015 catalog offers high precision photometric redshift measurements, stellar masses and star-formation rates. These estimates are derived through template fitting using the LEPHARE Spectral Energy Distribution (SED) fitting code \citep{Arnouts:2002, Ilbert:2006} with 30-band photometric data. The SED fitting procedure assumes a \citet{Chabrier:2003} Initial Mass Function (IMF), and delayed-$\tau$ star-formation histories ($\rm SFR \sim \tau^{-2}te^{-t/\tau}$ with e-folding time $\tau$). 

We obtain the $z$PDF and MASS$_{\rm MED}$ from the COSMOS2015 catalog \citep{Laigle:2016} as our photometric redshifts and stellar masses. In order to get objects with reliable photometric redshifts, we select objects labelled as galaxies ({\tt TYPE=0}) within the 2 deg$^2$ COSMOS area ({\tt FLAG\_COSMOS=1}) and the UltraVISTA area ({\tt FLAG\_HJMCC=0}), and not influenced by artifacts around saturated objects ({\tt FLAG\_PETER=0}). 

\citet{El-Badry:2016} analyzed 40 snapshots at $z<0.2$ for 8 simulated FIRE-2 galaxies and showed that the feedback-driven breathing mode is strongest in galaxies with $7.0<\rm \log M_*/M_\odot<9.6$ in the FIRE-2 simulations. We therefore select low-mass galaxies with $7.9<\rm \log M_*/M_\odot<9.6$ at $0.02<z<0.19$ from the COSMOS2015 catalog, in order to match the \citet{El-Badry:2016} sample while minimizing the impact of sample incompleteness given the limited depth of the observational data.

After applying these criteria, 2,012 galaxies are left in this sample. Note that the redshifts of 1,330 galaxies (43$\%$) in this sample have been confirmed by spectroscopic observations collected by the COSMOS team (Salvato et al., in prep).  Spectroscopic redshifts from several spectroscopic surveys are used when available, including spectroscopic redshifts from the $z$COSMOS survey \citep{Lilly:2007}, the VIMOS Ultra Deep Survey \citep[VUDS,][]{LeFevre:2015}, the Complete Calibration of the Color–Redshift Relation Survey \citep[C3R2,][]{Masters:2019}, the DEIMOS 10K Spectroscopic Survey \citep{Hasinger:2018}, and the FMOS-COSMOS Survey \citep{Kashino:2019}. 

\subsubsection{Keck/DEIMOS Spectroscopic data}

Our goal in this paper is to select a sample of dwarf galaxies to measure gas kinematics via emission lines with spectroscopic follow-up observations. This requires our dwarf galaxies to have ongoing star-formation in order to produce emission lines from photoionization. We therefore further select star-forming galaxies ({\tt CLASS=1}) in the COSMOS2015 catalog. The star-forming/quiescent classification flag is based on the best-fit absolute $\rm NUV-r$ and $\rm r-J$ colors from the 30-band SED fitting \citep{Laigle:2016}. Quiescent objects are defined by $\rm NUV-r>3(r-J)+1$ and $\rm NUV-r>3.1$.

We observed 127 star-forming dwarf galaxies with 13 slitmasks using the Deep Extragalactic Imaging Multi-Object Spectrograph \citep[DEIMOS,][]{Faber:2003} on the Keck-II telescope. Each slitmask was observed with one hour exposure and 1.0 arcsec slit width. We used the 1200G grating, which has a spectral resolution of R$\sim$4500 and covers a wavelength range of $\sim$ 2600\AA. We then reduced the DEIMOS data with PypeIt \citep{pypeit:zenodo,pypeit:joss_pub} and measured spectroscopic redshifts from the DEIMOS spectra. 

H$\alpha$ emission lines are detected in 103 star-forming dwarf galaxies. The top panel in Figure~\ref{ha_detection} shows $i$-band magnitude vs. $u-r$ color for all 127 observed low-mass galaxies with Keck/DEIMOS ($7.9<\rm \log M_*/M_\odot<9.6$ at $0.02<z<0.19$). Red points show galaxies with H$\alpha$ emission lines detected from our Keck/DEIMOS observations. Orange crosses show galaxies with other spectroscopic confirmations but without H$\alpha$ emission lines detected from our Keck/DEIMOS observations. Gray squares are galaxies without any successful spectroscopic confirmation, only 30-band photo-z's from the COSMOS2015 catalog. The middle and bottom panels show stellar masses vs. star-formation rates and specific star-formation rates for the same samples. Figure~\ref{completeness} shows the completeness of the H$\alpha$ detected galaxies relative to the photometric sample we selected from COSMOS2015, as a function of stellar masses (top), star-formation rates (middle), and specific star-formation rates (bottom). Error bars represent binomial uncertainties based on the number of galaxies in each bin. 

As shown in Figure~\ref{ha_detection} and Figure~\ref{completeness}, the completeness increases with stellar masses, star-formation rates and specific star-formation rates. Our H$\alpha$ detected sample (103 galaxies) is overall 81.1\% complete relative to the much deeper photometric sample selected from the COSMOS2015 catalog. 13 galaxies (10\%) have been spectroscopically confirmed by other observations (orange crosses in Figure~\ref{ha_detection}), but we do not detect usable H$\alpha$ emission lines to measure their velocity dispersions. They tend to be fainter and redder relative to the H$\alpha$ detected sample. They also tend to have lower SFR/sSFR and stellar masses. Their H$\alpha$ emissions may be too weak to be detected by our DEIMOS observations, or may have no H$\alpha$ emission at all. The 11 galaxies (8.3\%) without any spectroscopic confirmation (gray squares in Figure~\ref{ha_detection}) are even fainter and redder, with lower star-formation rates. They may have even weaker H$\alpha$ emissions. However, \citet{Laigle:2016} reported a catastrophic failure rate of $\sim$5\% for galaxies with $22<i<24$ in the COSMOS2015 catalog. Therefore, some of them could be galaxies at higher redshifts with no emission lines lie within our wavelength coverage. We also note that during the observation there was a CCD issue on Keck/DEIMOS. One of the CCD (CCD5) had a bias noise about 4 times higher than nominal, which decreased the S/N limit of the spectra on that CCD. This issue affected $\sim$25$\%$ of the area on DEIMOS masks. 

For the 103 H$\alpha$ detected dwarf galaxies, we measure the H$\alpha$ velocity dispersion and equivalent width with $\rm PPXF$ \citep{Cappellari:2023}. We corrected for the local velocity dispersion for instrumental dispersion with an average correction of 28 km/s. Note that the typical gas velocity dispersion for galaxies in our target mass range ($7.9<\rm \log M_*/M_\odot<9.6$) is 35 km/s. We calculate the uncertainties for the H$\alpha$ velocity dispersion based on the uncertainties from the fitting with $\rm PPXF$ and the DEIMOS instrumental broadening.

\begin{figure}
\includegraphics[width=8cm]{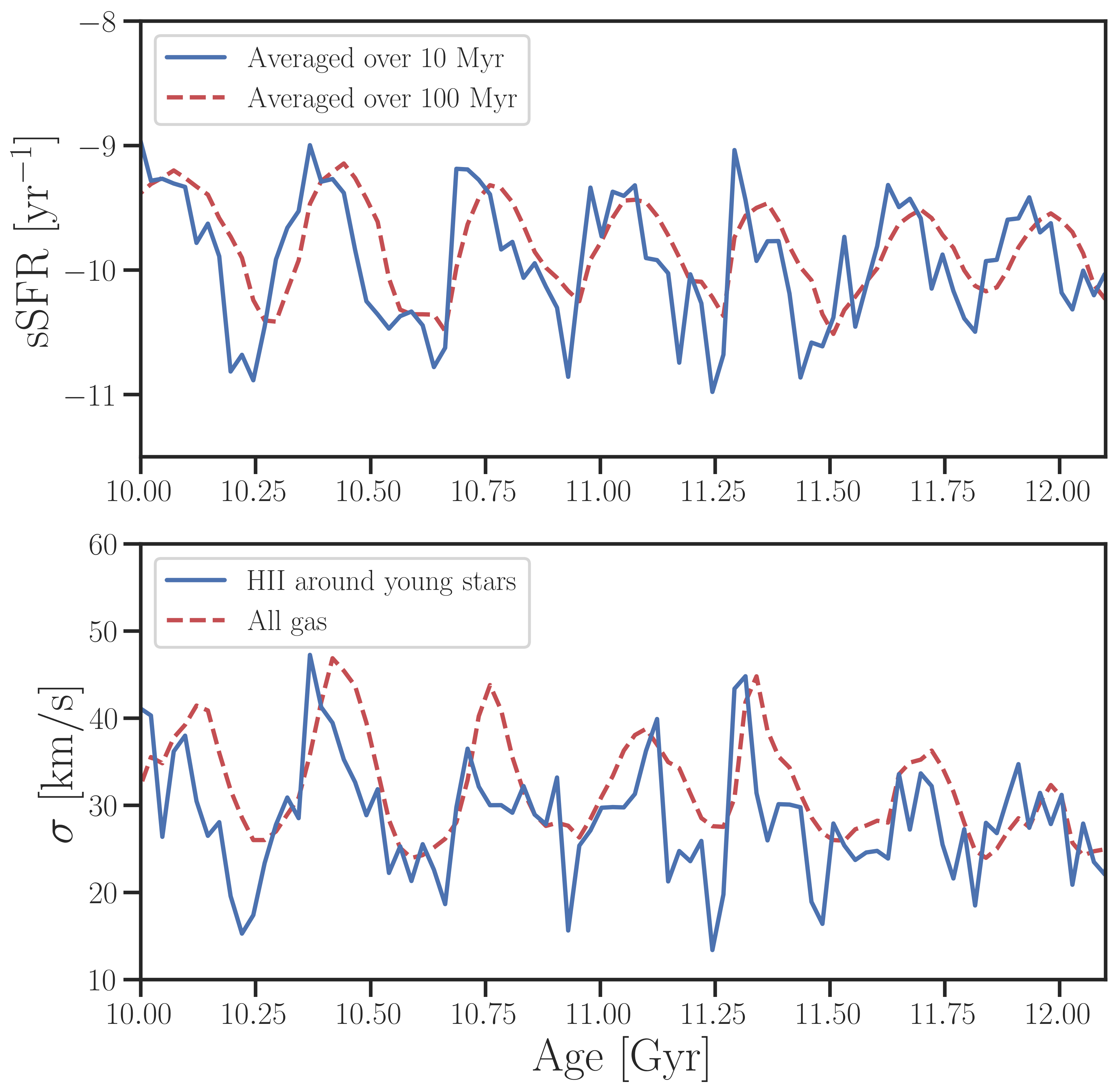}
\caption{Feedback-driven fluctuations in the FIRE-2 simulated galaxy $\rm m11d$ ($\rm \log M_*/M_\odot\sim9.6$). Top: sSFR averaged over 10 Myr (blue solid line, corresponding to sSFR derived from H$\alpha$) and 100 Myr (red dashed line, corresponding to sSFR derived from SED fitting). Bottom: Velocity dispersion measured with ionized HII gas around young stars (blue solid line, corresponding to $\rm \sigma_{gas}$ derived from H$\alpha$) and all gas (red dashed line). The breathing mode scenario suggests that gas responds directly to bursty star-formation, driving fluctuations in the gravitational potential and subsequently transferring the energy to stars and dark matter.
}
\label{FIRE_m11d}
\end{figure}

\begin{figure*}
\begin{center}
\includegraphics[width=16cm]{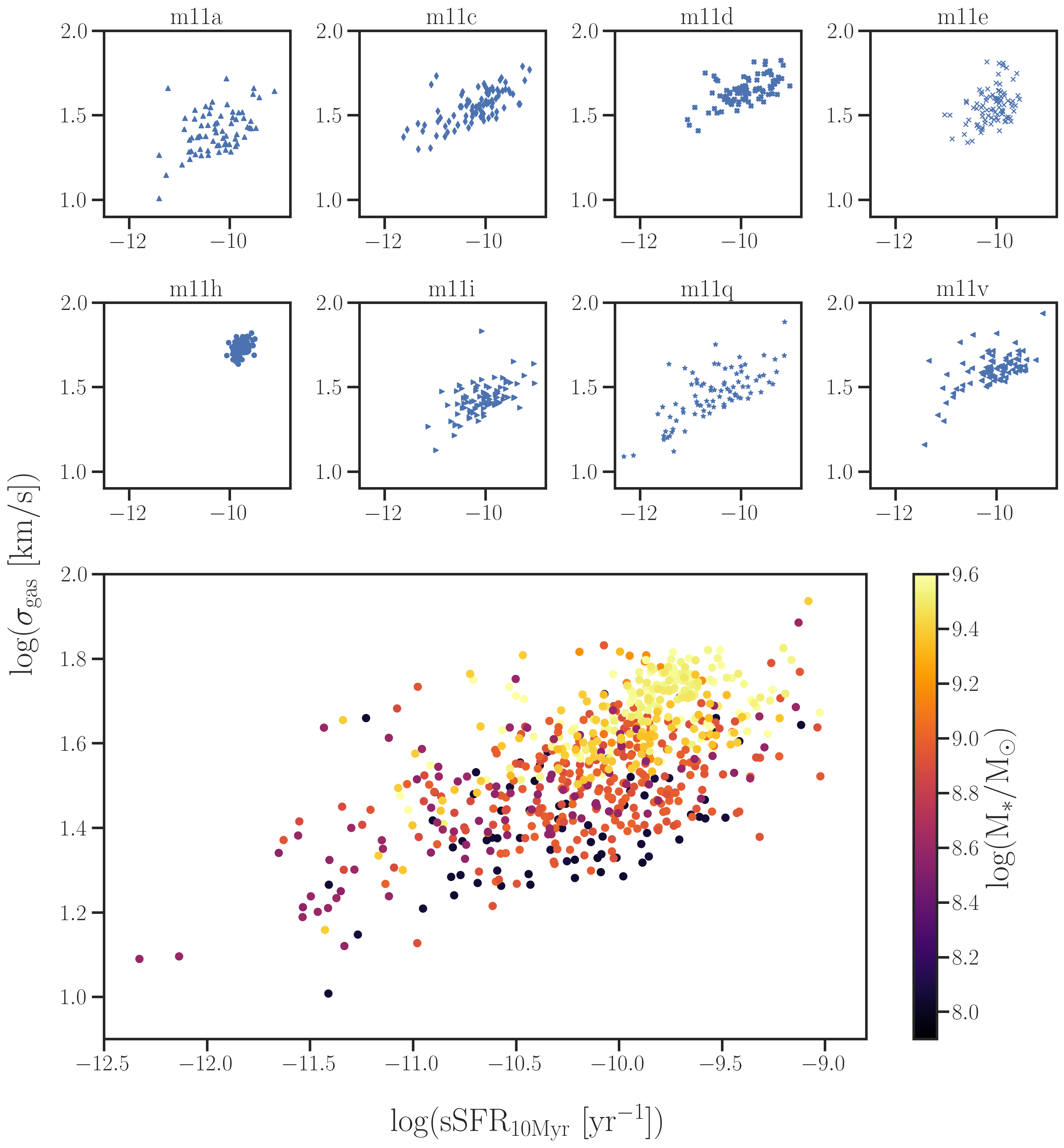}
\caption{Ionized gas velocity dispersions vs. specific star-formation rates averaged over 10 Myr for the low-mass galaxies in FIRE-2 simulations. We include 88 snapshots within $0.02<z<0.19$ for each FIRE-2 simulated galaxy. The upper 8 panels show $\rm \sigma_{gas}$ vs. $\rm sSFR_{10Myr}$ for $\rm m11a, m11c, m11d, m11e, m11h, m11i, m11q, m11v$ in FIRE-2, respectively. A positive relation between $\rm \sigma_{gas}$ and $\rm sSFR$, as explained by the feedback-driven breathing mode, can be found in all FIRE-2 simulated galaxies. Though this is weak for $\rm m11h$, which has a relatively less bursty SFH. This might be because that $\rm m11h$ is the diskiest galaxy among our sample, which acts to tamp down the burstiness of the SFH. The bottom panel show all the snapshots from the 8 simulated galaxies together, color-coded by the stellar mass. A clear mass dependence of $\rm \sigma_{gas}$ at a given sSFR can be seen.}
\label{FIRE_sigma_ssfr}
\end{center}
\end{figure*}

\subsubsection{Stellar masses and star-formation rates}

We re-run the SED fitting for our H$\alpha$ detected dwarf sample using \texttt{Prospector} \citep{Johnson:2021} with updated spectroscopic redshifts measured from Keck/DEIMOS. \texttt{Prospector} is a modular galaxy stellar populations inference code based on using stellar populations generated by the Flexible Stellar Population Synthesis (\texttt{FSPS}) code \citep{Conroy:2009,Conroy:2010}. We use the MIST stellar evolutionary isochrones \citep{Choi:2016}, the MILES spectral library \citep{Sanchez-Blazquez:2006}, the delayed-$\tau$ star-formation history with an additional instantaneous burst of star formation, the \citet{Chabrier:2003} Initial Mass Function, the \citet{Calzetti:2000} dust attenuation curve, and free gas-phase and stellar metallicities for the SED fitting. We also include nebular continuum and emission lines based on \texttt{CloudyFSPS} models \citet{Byler:2017}. We fit the SED with photometry from NUV to NIR obtained from the COSMOS2015 catalog \citep{Laigle:2016}, including GALEX-NUV, MegaCam-u$^*$, optical photometry from Subaru (including 6 broad-bands: B, V, g, r, i$^+$, z$^{++}$, 12 medium-bands: IA427, IA464, IA484, IA505, IA527, IA574, IA624, IA679, IA709, IA738, IA767, IA827, and 2 narrow-bands: NB711, NB816), HSC-y, and VIRCAM-Y/J/H/K$_s$. 

We obtain the best-fit stellar mass and sSFR for each galaxy from our SED fitting results. The sSFR is averaged over the last 100 Myr based on the best-fit star-formation histories (delayed-$\tau$ + an additional star-formation burst). Note that although the delayed-$\tau$ star-formation history has been used for many studies, it does not allow burstiness, which may be important for this work. Therefore we add one additional star-formation burst to the delayed-$\tau$ star-formation history in order to better estimate the star-formation rates within 100 Myr. 

We also derive the instantaneous sSFR ($\sim$10 Myr) of the dwarf galaxies with the H$\alpha$ equivalent width measured from the DEIMOS spectra, assuming there is no sSFR gradient. We first derive the H$\alpha$ luminosity $L_{\rm H\alpha}$ from the H$\alpha$ equivalent width and the integrated absolute r-band magnitude $M_r$:
\begin{equation}
L_{\mathrm{H}\alpha} = (\mathrm{EW} + \mathrm{EW}_c)\cdot 10^{-0.4(M_r - 34.1)}\times \frac{3 \times 10^{25} [\mathrm{erg\cdot s^{-1}}]}{[6564\cdot (1+z)]^2},
\end{equation}
where $\mathrm{EW}_c$ is a constant stellar absorption correction equals to 2.5\AA\ \citep{Gunawardhana:2011,Bauer:2013}. We then consider the internal dust corrections. Only a small subset of galaxies have H$\beta$ detection due to the short wavelength coverage of DEIMOS 1200G grating and insensitivity at bluer wavelengths. Therefore we take the best-fit dust extinction value from the \texttt{Prospector} SED fitting results described above, and convert it into the Balmer Decrement (BD) assuming the \citet{Calzetti:2000} dust attenuation curve. We then correct the H$\alpha$ luminosity $L_{\rm H\alpha}$:
\begin{equation}
L_{\mathrm{H}\alpha, \mathrm{corrected}} = L_{\mathrm{H}\alpha}\times (\frac{\mathrm{BD}}{2.86})^{2.36}.
\end{equation}
Finally, we determine instantaneous sSFR following \citet{Kennicutt:1998} and convert it into the \citet{Chabrier:2003} IMF by multiplying a factor of 0.63 \citep{Madau:2014}:
\begin{equation}
\mathrm{sSFR}\ [\mathrm{yr^{-1}}] = \frac{0.63\times 7.9\times 10^{-42}\cdot L_{\mathrm{H}\alpha, \mathrm{corrected}}}{\mathrm{M_{*}}\ [\mathrm{M_{\odot}}]},
\end{equation}
where $\rm M_{*}$ is the best-fit stellar mass from the \texttt{Prospector} SED fitting results.

Figure~\ref{example_galaxy} shows the SED of an example dwarf galaxy in our H$\alpha$ detected dwarf sample. The blue circles are the multi-band photometry obtained from the COSMOS2015 catalog and used for the SED fitting with \texttt{Prospector}. The black spectrum and gray boxes are the best-fit model spectrum and mock photometry from the SED fitting. The transmission curves of the filters used for the SED fitting are shown under the corresponding photometry. The lower two panels show the zoom-in regions around [OIII] and H$\alpha$. The lower left panel includes a color image of this galaxy from HSC PDR3. The red rectangle indicates the slit used for the Keck/DEIMOS observation. We also show the observed Keck/DEIMOS spectrum (red curves) for this object along with the model spectrum (black curves). Overall both the continuum and emission lines from the SED fitting agree well with the observation.

\begin{table*}
 \caption{Samples used in this paper.}
 \label{table:filter_info}
 \begin{tabular}{p{3.2cm}p{9.6cm}p{2.1cm}}
  \hline
  sample & description & sample size  \\
  \hline
  COSMOS2015 sample  & $0.02<z<0.19$, $7.9<\rm \log M_*/M_\odot<9.6$, photometric objects & 2012 galaxies\\
  H$\alpha$ detected sample & $0.02<z<0.19$, $7.9<\rm \log M_*/M_\odot<9.6$, star-forming, H$\alpha$ detected & 103 galaxies \\  
  FIRE-2 sample$^{\textrm{a}}$ & $0.02<z<0.19$, $7.9<\rm \log M_*/M_\odot<9.6$ & 704 snapshots\\
  \hline
  \end{tabular}\\
\small $^{\textrm{a}}$ We include 8 low-mass galaxies from the FIRE-2 simulations: $\rm m11a, m11c, m11d, m11e, m11h, m11i, m11q, m11v$. Each simulated galaxy has 88 snapshots between $0.02<z<0.19$. The typical time spacing between snapshots is 20 - 25 Myr.\\  
\label{sample_table}
\end{table*}

\subsection{FIRE-2 simulations}

To make a direct comparison between observations and simulations, we compute observables in the FIRE-2 simulations to match the observational data. We analyze 8 low-mass galaxies in the FIRE-2 simulations \citep{Hopkins:2018,Wetzel:2023}, including $\rm m11a, m11c, m11d, m11e, m11h, m11i, m11q, m11v$. These galaxies are cosmologically 'isolated', i.e. they are not in/near bigger groups/clusters. To match our observational sample, we use all snapshots, spaced typically by $\sim$25 Myr, across $0.02<z<0.19$ (88 snapshots for each galaxy, 864 snapshots in total). We compute the total stellar masses for all the snapshots. All 9 galaxies in the redshift range $0.02<z<0.19$ are in the same stellar mass range as our observational sample ($7.9<\rm \log M_*/M_\odot<9.6$). 

We identify young stars formed within the last 10 Myr/100 Myr and compute the specific star-formation rates (sSFR) averaged over the last 10 Myr/100 Myr, in order to match the sSFR derived from the SED fitting and H$\alpha$ emission lines \citep{Jose:2021}. We then compute the mass-weighted velocity dispersion $\rm \sigma_{gas}$ for the ionized gas around HII regions. For each snapshot, we identify ionized gas particles within 100 pc of young stars (age less than 10Myr), to match the observed $\rm H\alpha$ emissions which are ionized from the O/B stars in HII regions. We also compute the $\rm \sigma_{gas}$ for all ionized gas (not only around young stars) in FIRE-2 simulations. The main results in the following sections will not change when using $\rm \sigma_{gas}$ for all ionized gas. In order to match the observed $\rm H\alpha$ emissions, we will use the $\rm \sigma_{gas}$ around the young stars for further analysis.

To match the inclination effect in observations, we choose 100 random line-of-sights and take the median value of the dispersion. However, the inclination effect itself could contribute to the scatter in $\rm \sigma_{gas}$ in a given snapshot. We therefore compute the 1-$\sigma$ scatter of the $\rm \sigma_{gas}$ over the 100 random line-of-sights in every snapshot to be the uncertainty of the $\rm \sigma_{gas}$ in the simulation. 

Figure~\ref{FIRE_m11d} shows the feedback-driven fluctuations in one FIRE-2 simulated galaxy in our sample $\rm m11d$ ($\rm \log M_*/M_\odot\sim9.6$) as an example. The top panel shows the fluctuations in sSFR over 10 Myr (blue solid line) and 100 Myr (red dashed line). The bottom panel shows the fluctuations in velocity dispersion measured with ionized HII gas around young stars (blue solid line) and all gas (red dashed line). This work aims to test the short-timescale dynamical effect driven by the breathing mode, i.e. the correlation between sSFR and $\rm \sigma_{gas}$ across a single starburst episode.

All samples presented in this paper from both observations and simulations are summarized in Table~\ref{sample_table}.

\begin{figure*}
\begin{center}
\includegraphics[width=16cm]{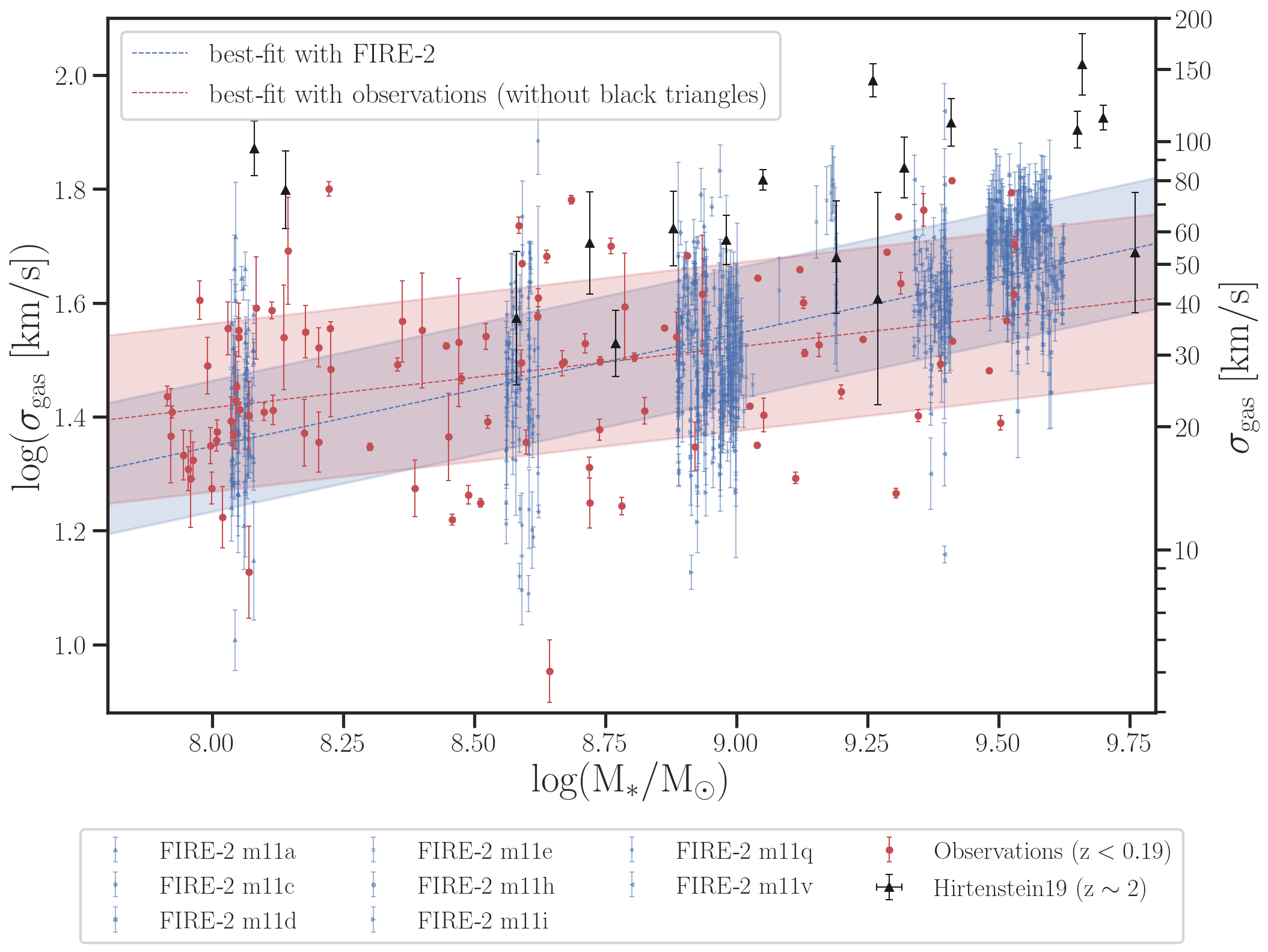}
\caption{The ionized gas velocity dispersion vs. stellar mass for the low-mass galaxies. Red points with error bars are the H$\alpha$ detected dwarf sample at $z<0.19$. Blue data points with different symbols show the 8 simulated low-mass galaxies (88 snapshots within $0.02<z<0.19$ for each simulation) in FIRE-2 simulations. Error bars for the simulated galaxies indicate the 1-$\sigma$ scatter of the gas velocity dispersion measured from 100 random line-of-sights in each FIRE-2 snapshot. The blue dashed line shows the best-fit relation for the simulation data. Black triangles show the observed low-mass galaxies at $z\sim2$ in \citet{Hirtenstein:2019}. Red and blue shaded areas show the 1$\sigma$ scatter (68\%) around the best-fit relation (dashed lines) for the H$\alpha$ detected dwarf sample and the FIRE-2 sample. The best-fit $\rm \sigma_{gas}$-mass relations from both FIRE-2 simulations and our observations match within 1$\sigma$. The $z\sim2$ sample from \citet{Hirtenstein:2019} lies above the relation due to their systematically higher sSFR.}
\label{vdisp_mass}
\end{center}
\end{figure*}

\section{Results}\label{results}

In this section we present the gas kinematics and star-formation activities in the observed low-mass galaxies, and compare it with FIRE-2 simulations. 

\subsection{The mass dependence in $\rm \sigma_{gas}$-sSFR relation}\label{sigma_gas}

The feedback-driven breathing mode predicts a correlation between gas velocity dispersion and star-formation activities \citep{El-Badry:2016,El-Badry:2017,Hirtenstein:2019}. We show this relation again with 8 low-mass galaxies with stellar masses between $7.9<\rm \log M_*/M_\odot<9.6$ from the FIRE-2 simulations in Figure~\ref{FIRE_sigma_ssfr}. The upper 8 panels show ionized gas velocity dispersion $\rm \sigma_{gas}$ vs. sSFR averaged over 10 Myr for the 88 snapshots between $0.02<z<0.19$ in FIRE-2-$\rm m11a, m11c, m11d, m11e, m11h, m11i, m11q, m11v$, respectively. As can be seen, all of the 8 FIRE-2 low-mass galaxies show positive $\rm \sigma_{gas}$-sSFR relations, though it is weak for $\rm m11h$. This might be because $\rm m11h$ is the diskiest galaxy among our sample, which acts to tamp down the burstiness of the SFH. In the lower panel of Figure~\ref{FIRE_sigma_ssfr} we show all 8 FIRE-2 simulated galaxies together. A clear mass dependence in $\rm \sigma_{gas}$ can be seen, i.e. at given sSFR more massive galaxies have higher $\rm \sigma_{gas}$. Therefore, in order to test the relation between gas velocity dispersion and star-formation activities in galaxies with different masses, we first need to correct for the mass dependence of $\rm \sigma_{gas}$. 

Figure~\ref{vdisp_mass} shows the gas velocity dispersion vs. stellar mass for three different samples. Red points with error bars are the H$\alpha$ detected dwarf sample. Blue points with different symbols are the 8 low-mass galaxies in FIRE-2 simulations. Black triangles are the observed low-mass galaxies at $z\sim2$ from \citet{Hirtenstein:2019}. Error bars for the observed samples are measurement uncertainties. For the FIRE-2 sample, the error bars are the 1-$\sigma$ scatter of $\rm \sigma_{gas}$ over the 100 random line-of-sights, representing the inclination effect in each snapshot. We show the best-fit and 1-$\sigma$ scatter around the best-fit for both the FIRE-2 sample (blue dashed line and blue shaded area) and the H$\alpha$ detected dwarf sample (red dashed line and blue shaded area).

Overall, galaxies in the H$\alpha$ detected dwarf sample and FIRE-2 sample occupy a similar range of $\rm \sigma_{gas}$ at a given stellar mass. There is a positive mass-$\rm \sigma_{gas}$ relation in both data sets. The \citet{Hirtenstein:2019} sample overall has higher $\rm \sigma_{gas}$ which can be attributed to higher sSFR in this sample, but still follows the same positive trend. The best-fit mass-$\rm \sigma_{gas}$ relations from both FIRE-2 simulations and our observations lie within the 1-$\sigma$ region of each other. Because our H$\alpha$ detected dwarf sample is also 10-20\% incomplete at the low-mass/low-SFR end, the slope of the $\rm \sigma_{gas}$-mass relation could be affected. Therefore we choose to use the $\rm \sigma_{gas}$-mass relation in the FIRE-2 sample ($\rm log\sigma_{gas} = 0.198\times logM_{*}/Msun - 0.235$) and compute the residual value $\rm \Delta\sigma_{gas}$ with the same formula for all the observed and simulated galaxies, i.e. 
\begin{equation}
    \rm \Delta\sigma_{gas} = log\sigma_{gas} - (0.198\times logM_{*}/Msun - 0.235).
\end{equation}
We also show the results using a different mass-$\rm \sigma_{gas}$ relation in Appendix~\ref{diffslopes}. We show that the main results do not change when using the observed mass-$\rm \sigma_{gas}$ relation for the observed sample.

\subsection{$\rm \Delta\sigma_{gas}$ vs. specific star-formation rates}\label{sigma_mass}

\begin{figure*}
\begin{center}
\includegraphics[width=15cm]{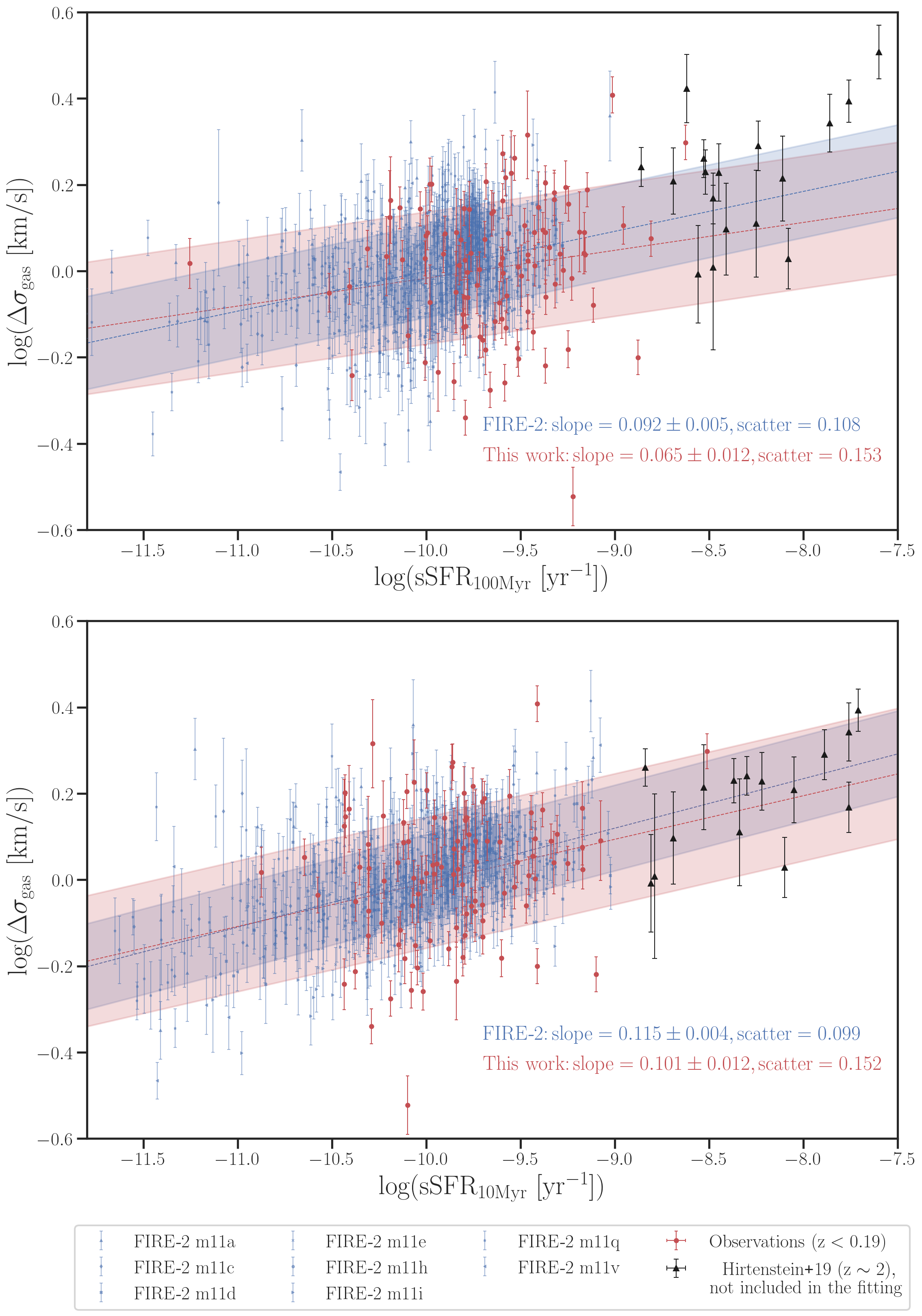}
\caption{$\rm \Delta\sigma_{gas}$ vs. sSFR averaged over 100 Myr (upper panel) and 10 Myr (lower panel) for low-mass galaxies from both observations and simulations. Red points with error bars are the H$\alpha$ detected dwarf sample at $z<0.19$. Blue data points with different symbols show the 8 simulated low-mass galaxies (88 snapshots within $0.02<z<0.19$ for each simulation) in FIRE-2 simulations. Error bars for the simulated galaxies indicate the 1-$\sigma$ scatter of the gas velocity dispersion measured from 100 random line-of-sights in each FIRE-2 snapshot. Black triangles are the observed low-mass galaxies at $z\sim2$ in \citet{Hirtenstein:2019}, which show a similar trend extending to higher sSFR. Red and blue shaded areas show the 1$\sigma$ scatter (68\%) around the best-fit relation (dashed lines) for the H$\alpha$ detected dwarf sample and the FIRE-2 sample.}
\label{dsigma_ssfr}
\end{center}
\end{figure*}

We now present the relation between $\rm \Delta\sigma_{gas}$ and specific star-formation rates in our galaxies. We fit a linear relation between $\rm \Delta\sigma_{gas}$ and sSFR for the FIRE-2 sample and the H$\alpha$ detected dwarf sample, weighted by their uncertainties. The upper panel in Figure~\ref{dsigma_ssfr} shows the $\rm \Delta\sigma_{gas}$ vs. sSFR averaged over the last 100 Myr for three samples. Red and blue points with error bars are the H$\alpha$ detected dwarf sample and the FIRE-2 sample. Black triangles show the observed low-mass galaxies at $z\sim2$ from \citet{Hirtenstein:2019}. We also show the 1$\sigma$ scatter regions (68\%) around the best-fit relations (dashed lines) for the H$\alpha$ detected dwarf sample (red shaded area) and the FIRE-2 sample (blue shaded area).

Both the FIRE-2 sample and the H$\alpha$ detected dwarf sample show a positive $\rm \Delta\sigma_{gas}$-$\rm sSFR_{100Myr}$ relation. The best-fit lines of both samples lie within the 1$\sigma$ regions of the other sample in the plot. The 1$\sigma$ scatter for the FIRE-2 sample (0.108 dex) is smaller than the H$\alpha$ detected dwarf sample (0.153 dex). However this might not be surprising, given that the observed sample contains ~100 galaxies, while our simulated sample is a time-sample of only 8 galaxies. The relation for the FIRE-2 sample (slope=0.092) is also steeper than the H$\alpha$ detected dwarf sample (slope=0.065). The $z\sim2$ sample from \citet{Hirtenstein:2019} is clearly biased to higher sSFR because of the selection effects. Nevertheless, they are on the same positive trend of the $\rm \Delta\sigma_{gas}$-$\rm sSFR_{100Myr}$ relation. Over 88\% (15/17) of the galaxies in the $z\sim2$ sample lie within the 1$\sigma$ scatter around the best-fit lines of either the FIRE-2 sample or the H$\alpha$ detected dwarf sample.

The lower panel in Figure~\ref{dsigma_ssfr} shows the same plot as the upper panel, except for using sSFR averaged over the last 10 Myr along the x-axis. Both the FIRE-2 sample and the H$\alpha$ detected dwarf sample show a similar positive $\rm \Delta\sigma_{gas}$-sSFR relation. The slopes of the $\rm \Delta\sigma_{gas}$-$\rm sSFR_{10Myr}$ relation for both samples are steeper than the $\rm \Delta\sigma_{gas}$-$\rm sSFR_{100Myr}$ relation. For the $\rm \Delta\sigma_{gas}$-$\rm sSFR_{10Myr}$ relation, the FIRE-2 sample (slope=0.115) also shows a steeper relation than the H$\alpha$ detected dwarf sample (slope=0.101), similar to the $\rm \Delta\sigma_{gas}$-$\rm sSFR_{100Myr}$ relation.

For the FIRE-2 sample, the 1$\sigma$ scatter (0.099 dex)  in the  $\rm \Delta\sigma_{gas}$-$\rm sSFR_{10Myr}$ relation is even smaller than the 1$\sigma$ scatter (0.152 dex) in the $\rm \Delta\sigma_{gas}$-$\rm sSFR_{100Myr}$ relation. In both plots, the observed galaxies show larger scatter than the FIRE-2 simulations. We note again that this might be due to the difference in the sample size between the observational sample and the FIRE-2 sample. The $z\sim2$ galaxies in \citet{Hirtenstein:2019} also follow the same $\rm \Delta\sigma_{gas}$-$\rm sSFR_{10Myr}$ relation. All of the 17 galaxies in the $z\sim2$ sample lie within the 1$\sigma$ scatter around the best-fit lines of either the FIRE-2 sample or the H$\alpha$ detected dwarf sample.

We also show the uncertainties of the slope in the $\rm \Delta\sigma_{gas}$-$\rm sSFR$ relations. The FIRE-2 sample has smaller uncertainties for both $\rm \Delta\sigma_{gas}$-$\rm sSFR_{100Myr}$ (0.005) and $\rm \Delta\sigma_{gas}$-$\rm sSFR_{10Myr}$ (0.004), while the observed sample has larger uncertainties (0.012 for both $\rm \Delta\sigma_{gas}$-$\rm sSFR_{100Myr}$ and $\rm \Delta\sigma_{gas}$-$\rm sSFR_{10Myr}$, respectively). Nevertheless, both simulations and observations show clear positive/non-zero slopes between $\rm \Delta\sigma_{gas}$ and sSFR.

We also highlight that in Figure~\ref{dsigma_ssfr}, our H$\alpha$ detected dwarf sample has very few points below $\rm log(sSFR)<-10.5\ yr^{-1}$. However, the FIRE-2 sample has much more points below $\rm sSFR<-10.5\ yr^{-1}$. This could be due to our selection effect, since we need the galaxies to have enough recent star-formation to produce H$\alpha$ emission lines, in order to measure the $\rm \sigma_{gas}$. Or it could be due to the unique condition of the FIRE-2 simulations we use. 

\begin{figure*}
\begin{center}
\includegraphics[width=15cm]{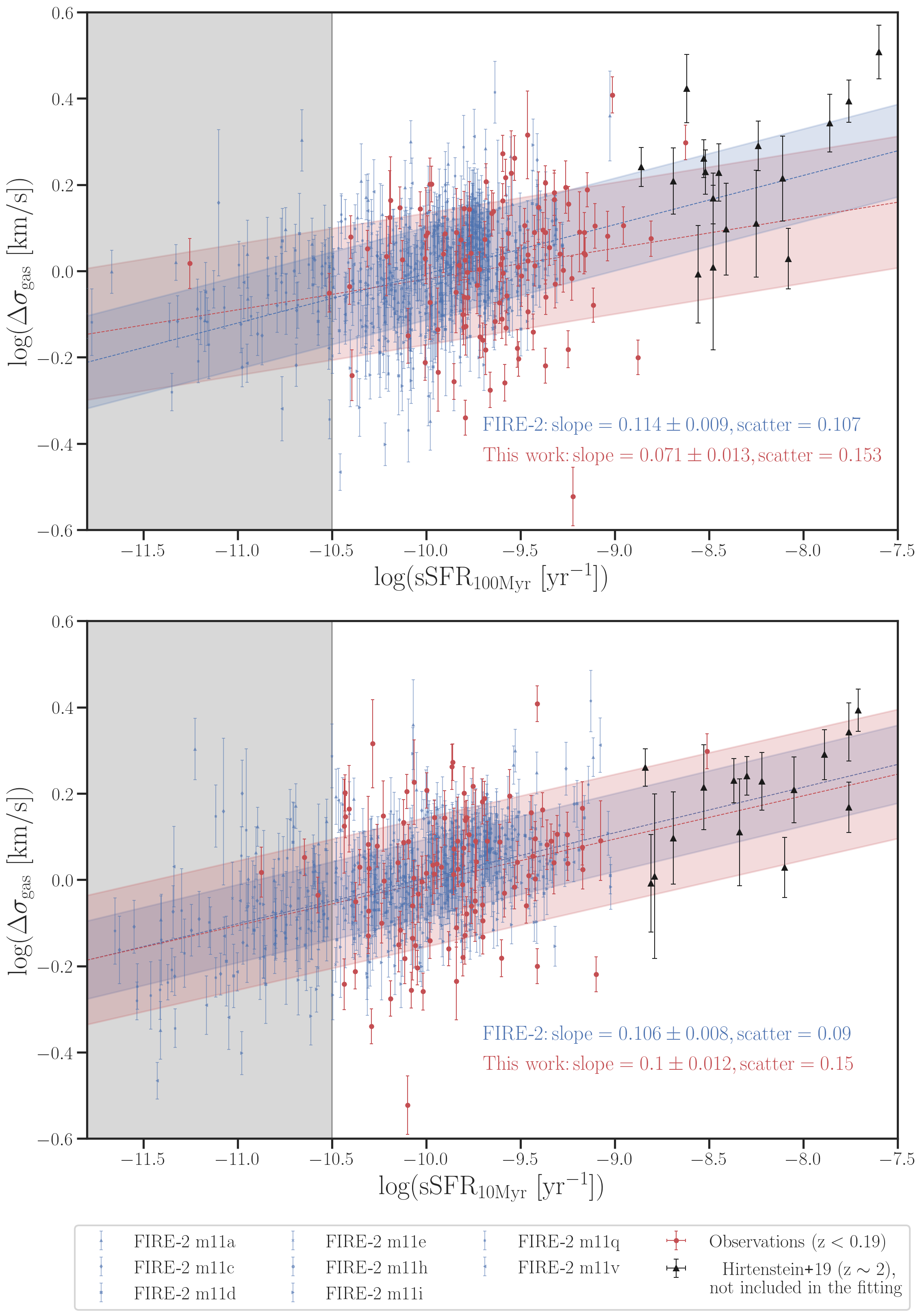}
\caption{Same as Figure~\ref{dsigma_ssfr}, but excluding galaxies with $\rm log(sSFR)<-10.5\ yr^{-1}$ (shaded regions) from the fit. The FIRE-2 simulations and our observations still show similar $\rm \Delta\sigma_{gas}$–$\rm sSFR$ relations for both sSFR indicators, although the relations from FIRE-2 are tighter. Even with this more star-forming sub-sample, the $\rm \Delta\sigma_{gas}$–$\rm sSFR$ relation remains largely unchanged compared to Figure~\ref{dsigma_ssfr}. Therefore, the lack of lower-sSFR galaxies in the H$\alpha$ detected sample does not affect the main conclusions.}
\label{dsigma_ssfr_cut}
\end{center}
\end{figure*}

In order to make a fairer comparison, we drop all the galaxies with $\rm log(sSFR)<-10.5\ yr^{-1}$ in both our observations and FIRE-2 simulations and re-fit the $\rm \Delta\sigma_{gas}$-$\rm sSFR$ relations. Figure~\ref{dsigma_ssfr_cut} shows the relations with this $\rm log(sSFR)>-10.5\ yr^{-1}$ sample. The shaded area indicates the low SSFR region that we do not use for the new fitting, though we still show the points in the figure. Similar to Figure~\ref{dsigma_ssfr}, with this more star-forming sub-sample, the $\rm \Delta\sigma_{gas}$-$\rm sSFR$ relations do not change much. The $\rm \Delta\sigma_{gas}$-$\rm sSFR_{100Myr}$ relation in observations (slope=0.071) is still shallower than the FIRE-2 simulations (slope=0.114). The $\rm \Delta\sigma_{gas}$-$\rm sSFR_{10Myr}$ relation from observations (slope=0.1) is still very close to FIRE-2 simulations (slope=0.106). The scatters from FIRE-2 simulations (0.107/0.09 dex) are smaller than the observations (0.153/0.15 dex), similar to what we have seen in Figure~\ref{dsigma_ssfr}. Therefore, the fact that our observational sample lacks low-sSFR galaxies does not change the main results in the $\rm \Delta\sigma_{gas}$-$\rm sSFR$.

\begin{figure*}
\begin{center}
\includegraphics[width=16cm]{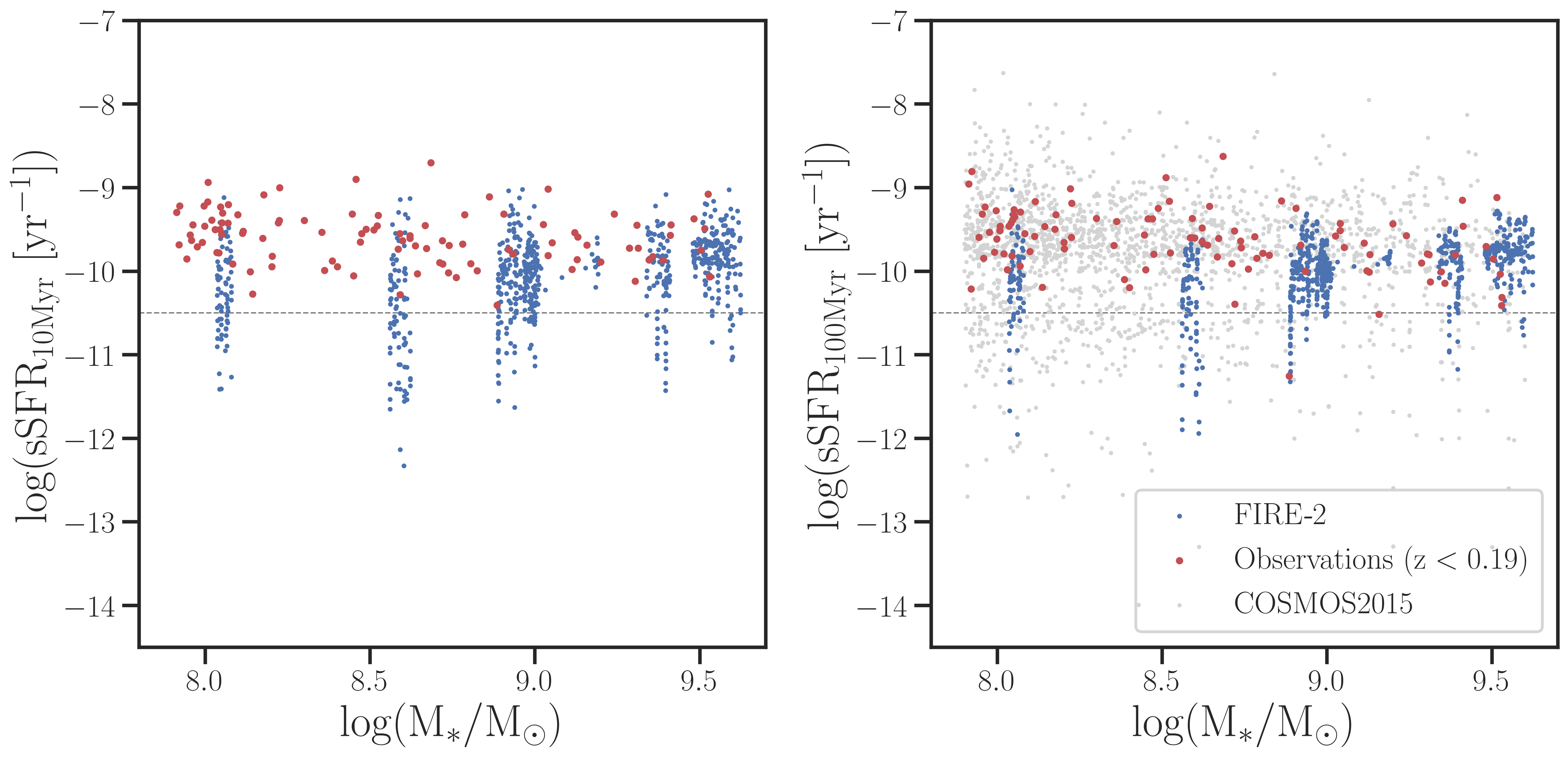}
\caption{Left panel: stellar mass vs. sSFR (10 Myr) for the $\rm H\alpha$ detected sample at $z<0.19$ (red points) and the 8 low-mass galaxies in FIRE-2 simulations (blue points). Right panel: stellar mass vs. sSFR (100 Myr) for the $\rm H\alpha$ detected sample at $z<0.19$ (red points) and all low-mass galaxies in the COSMOS2015 catalog with spec-$z$'s and only photo-$z$'s (gray points). The gray dashed lines show the sSFR limit ($\rm -10.5\ yr^{-1}$) used in Figure~\ref{dsigma_ssfr_cut}.}
\label{ssfr_mass}
\end{center}
\end{figure*}

\subsection{The distribution of sSFR in low-mass galaxies}\label{ssfr}

We now turn to the topic of the distribution of sSFR in the mass range $7.9<\rm \log M_*/M_\odot<9.6$. 

The left panel in Figure~\ref{ssfr_mass} shows the $\rm sSFR_{10Myr}$ vs. stellar mass for the FIRE-2 sample (blue points) and our H$\alpha$ detected dwarf sample (red points). The right panel in Figure~\ref{ssfr_mass} shows the same plot except for using $\rm sSFR_{100Myr}$ as the y-axis. The H$\alpha$ detected dwarf sample in both panels on average has clearly higher sSFR than the FIRE-2 sample at nearly all masses. 

We include the dwarf COSMOS2015 sample (gray points) in the right panel in Figure~\ref{ssfr_mass}. The stellar masses and $\rm sSFR_{100Myr}$ in the COSMOS2015 dwarf sample are based on the 30-band SED fitting in the COSMOS2015 catalog \citep{Laigle:2016}, which is much deeper than our H$\alpha$ detected dwarf sample. As shown in the right panel in Figure~\ref{ssfr_mass}, the H$\alpha$ detected dwarf sample generally follows the star-forming main sequence indicated by the COSMOS2015 dwarf sample. 

\begin{figure*}
\begin{center}
\includegraphics[width=16cm]{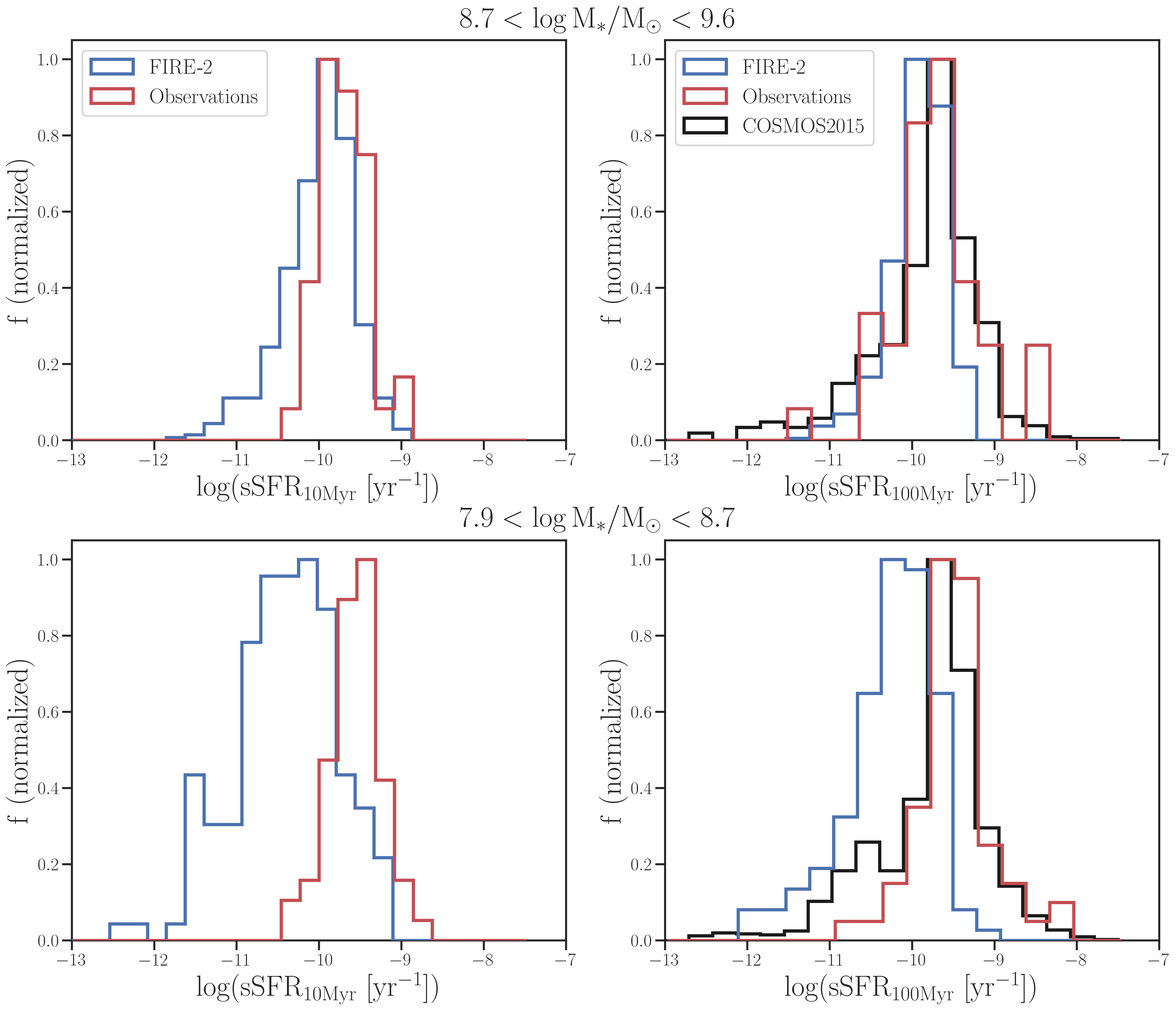}
\caption{Normalized histograms of sSFR for low-mass galaxies from observations and simulations across two stellar mass bins: $8.7<\rm \log M_*/M_\odot<9.6$ (upper panels) and $7.9<\rm \log M_*/M_\odot<8.7$ (lower panels). Left panels show sSFRs averaged over 10 Myr for the FIRE-2 sample (blue histogram) and the H$\alpha$ detected dwarf sample (red histogram). Right panels show the distributions of $\rm sSFR_{100Myr}$ for the FIRE-2 sample (blue histogram), the H$\alpha$ detected dwarf sample (red histogram), and the COSMOS2015 dwarf sample (black histogram).}
\label{ssfr_hist}
\end{center}
\end{figure*}

Figure~\ref{ssfr_hist} shows the distributions of $\rm sSFR_{10Myr}$ (left panels) and $\rm sSFR_{100Myr}$ (right panels) for two mass bins: $8.7<\rm \log M_*/M_\odot<9.6$ (upper panels) and $7.9<\rm \log M_*/M_\odot<8.7$ (lower panels). In the left panels, we show the distributions of $\rm sSFR_{10Myr}$ for the FIRE-2 sample (blue histogram) and the H$\alpha$ detected dwarf sample (red histogram). In the right panels, we show the distributions of $\rm sSFR_{100Myr}$ for the FIRE-2 sample (blue histogram), the H$\alpha$ detected dwarf sample (red histogram), and the COSMOS2015 dwarf sample (black histogram). 

For the higher mass bin ($8.7<\rm \log M_*/M_\odot<9.6$), the $\rm sSFR_{100Myr}$ distributions from all three samples are similar. The only difference is $\rm sSFR_{10Myr}$ between the FIRE-2 sample and the H$\alpha$ detected dwarf sample, i.e., the FIRE-2 sample has a small fraction of galaxies below $\rm log(sSFR_{10Myr})<-10.5\ yr^{-1}$ while the H$\alpha$ detected dwarf sample does not. This discrepancy is more significant for the lower mass bin ($7.9<\rm \log M_*/M_\odot<8.7$). Our H$\alpha$ detected dwarf sample does not have many galaxies with $\rm log(sSFR)<-10.5\ yr^{-1}$, while the FIRE-2 sample has a significant fraction of galaxies with $\rm log(sSFR)<-10.5\ yr^{-1}$, for both $\rm sSFR_{100Myr}$ and $\rm sSFR_{10Myr}$.

However, for the deeper and more complete sample from COSMOS2015, it does show a tail extended to very low sSFR of $\rm log(sSFR_{100Myr})\sim -13.0\ yr^{-1}$, which covers the low full dynamic range of $\rm sSFR_{100Myr}$ in our FIRE-2 sample. We also note that both observational samples also have high sSFR objects ($\rm log(sSFR_{100Myr})>-9.0\ yr^{-1}$) that the FIRE-2 simulations do not produce. This high sSFR tail can also be seen in our H$\alpha$ detected dwarf sample. We note that the $\rm log(sSFR_{100Myr}$ used for the FIRE-2 sample are computed directly based on the young stars in the simulations, not exactly the same as the SED modeling using observational data. This may be improved by fitting the mock SED of the FIRE-2 sample using synthetic photometry in the same filters as the observations. We leave this to future studies.

\section{Discussion}\label{discussion}

\subsection{Velocity dispersion of ionized gas}

Our first finding is that FIRE-2 simulations produce a realistic range of $\rm \sigma_{gas}$, compared to the $\rm \sigma_{gas}$ indicated by H$\alpha$ in observed low-mass galaxies. Our observed low-mass galaxies generally follow the same $\rm \sigma_{gas}$-mass relation from the FIRE-2 simulations, with a comparable scatter in $\rm \sigma_{gas}$ at a given stellar mass. For galaxies with $\rm 7.9<\log M_*/M_\odot<9.6$, the FIRE-2 simulations predict a mean $\rm \sigma_{gas}$ of 37.4 km/s with a standard deviation of 12.2 km/s, while our observations show a mean $\rm \sigma_{gas}$ of 36.1 km/s with a standard deviation of 11.6 km/s. The range of $\rm \sigma_{gas}$ in this mass range, particularly at the higher mass end (e.g. $\rm 9.0<\log M_*/M_\odot<9.6$), also agrees with the range of $\rm \sigma_{gas}$ measured from H$\alpha$ in previous studies \citep[][]{Green:2014,Johnson:2018,Yu:2019}. The galaxies in the $z\sim 2$ sample from \citet{Hirtenstein:2019} lie at the upper boundary of the $\rm \sigma_{gas}$-mass relation indicated by FIRE-2 simulations since this sample has higher sSFR, which is expected to have high $\rm \sigma_{gas}$ at given a stellar mass according to the breathing mode scenario. 

Although the $\rm \sigma_{gas}$ distribution at a given stellar mass in our observational sample generally agrees with the FIRE-2 simulations, it shows a shallower best-fit mass-$\rm \sigma_{gas}$ relation than FIRE-2. This could be due to the selection effect, i.e. the H$\alpha$ detected dwarf sample, is $\sim$ 20\% incomplete for low mass and low sSFR galaxies. According to the breathing mode scenario, those low $\rm \sigma_{gas}$ galaxies are at the low sSFR phase \citep{El-Badry:2017}, which means they also have weaker H$\alpha$ emission lines. The combination of the low mass (i.e. fainter) and weak H$\alpha$ make those galaxies very hard to detect, even if they exist. Another possible explanation is due to the spectral resolution of our observation. The mean $\rm \sigma_{gas}$ at $\rm \log M_*/M_\odot\sim 7.9$ predicted by the $\rm \sigma_{gas}$-mass relation in FIRE-2 is 21.3 km/s, while the instrumental dispersion $\rm \sigma_{inst}$ of our Keck/DEIMOS spectra is $\sim$28 km/s. The instrumental dispersion might not allow us to properly measure the velocity dispersion for lower $\rm \sigma_{gas}$. Another caveat is that, while we use 1000's of snapshots, the simulations only have 8 independent galaxies, and a larger sample of simulations would help bring the statistical significance of independent formation histories more similar to the observational sample of ~100 galaxies.

\subsection{Feedback-driven breathing mode in low-mass galaxies: observations vs. simulations}

A key prediction of the feedback-driven breathing mode in low-mass galaxies is the relation between the gas kinematics and recent star-formation activities, i.e. at a given stellar mass, higher SFR results in higher $\rm \sigma_{gas}$ \citep{El-Badry:2017, Hirtenstein:2019}. With our observations on Keck/DEIMOS ($\rm \sigma_{inst}\sim 28\ km/s$ ), we confirm the positive $\rm \Delta\sigma_{gas}$-sSFR relation predicted by the FIRE-2 simulations in low-mass galaxies ($\rm 7.9<\log M_*/M_\odot<9.6$). As shown in Figure~\ref{dsigma_ssfr}, $\rm \Delta\sigma_{gas}$ has positive correlations with two independent sSFR indicators: $\rm sSFR_{100Myr}$, which is derived from SED fitting; and $\rm sSFR_{10Myr}$, which is based on $\rm H\alpha$. For both observations and simulations, the best-fit $\rm \Delta\sigma_{gas}$-sSFR relations (blue and red dashed lines in Figure~\ref{dsigma_ssfr}) lie within the 1-$\sigma$ scatter region of each other (blue and red shaded areas in Figure~\ref{dsigma_ssfr}). 

We notice that the observed best-fit slope is shallower than the simulations for $\rm sSFR_{100Myr}$, while $\rm sSFR_{10Myr}$ shows a closer slope to the simulations. The observations have larger 1-$\sigma$ scatter than the simulations. If we assume that the statistical uncertainty from observations is well characterized by the error bars and the fitting of the relation presented in Figure~\ref{dsigma_ssfr}, then although the positive $\rm \Delta\sigma_{gas}$-sSFR relation is observed, it might be less significant than the prediction from FIRE-2 simulations. In the FIRE-2 simulations, the $\rm \Delta\sigma_{gas}$-sSFR relation is produced by the feedback-driven breathing mode, i.e. stellar feedback and bursty star-formation activities. The significance of the $\rm \Delta\sigma_{gas}$-sSFR relation in simulations is governed by the strength of the stellar feedback-driven potential fluctuations, and the burstiness of the star-formation histories \citep{El-Badry:2016, El-Badry:2017}. The combination of strong and repeated feedback-driven potential fluctuations would produce a tight $\rm \Delta\sigma_{gas}$-sSFR relation. The slight difference between observations and simulations in the $\rm \Delta\sigma_{gas}$-sSFR relation could be due to the difference in the strength of the feedback and the burstiness of the star-formation history, or the limited sample of individual FIRE-2 simulated galaxies. In addition, the $\rm sSFR_{100Myr}$ for our observational sample is derived from 30-band SED fitting using the delayed-$\tau$ star-formation history. This parametric star-formation history does not allow the bursty star-formation in its model. Therefore, the $\rm sSFR_{100Myr}$ may not be able to capture the recent bursty star-formation that are producing the feedback in our low-mass galaxies. The fact that the observed $\rm \Delta\sigma_{gas}$-sSFR relation is shallower when using $\rm sSFR_{100Myr}$ may be due to that $\rm sSFR_{100Myr}$ from SED fitting is not flexible enough, while $\rm sSFR_{10Myr}$ derived directly from H$\alpha$ is sensitive to bursty star-formation activities and does not depend on any SFH assumptions. Using more flexible star-formation history models (e.g. non-parametric star-formation histories) may improve the sSFR measurements, while also yielding more accurate stellar mass estimates \citep[e.g.][]{Leja:2017, Iyer:2019, Leja:2019, Li:2022, delosReyes:2025}.

Another difference between our observations and FIRE-2 simulations that can be seen in Figure~\ref{dsigma_ssfr} (also in Figure~\ref{ssfr_mass} and Figure~\ref{ssfr_hist}) is the lack of low sSFR galaxies in our observational sample. The difference is more significant in the lower mass bin ($\rm 7.9<\log M_*/M_\odot<8.7$) than the higher mass bin ($\rm 8.7<\log M_*/M_\odot<9.6$). The sample incompleteness could explain this, since in order to measure the gas kinematics from H$\alpha$, our observed galaxies need to have recent star-formation to ionize the HII regions in the galaxies. Therefore our sample by definition does not have momentarily quiescent objects, i.e. galaxies with very low short-timescale $\rm sSFR$. Further observations with stellar kinematics (e.g. using $\sigma_{\rm star}$ instead of $\sigma_{\rm gas}$ from H$\alpha$) could allow us to probe the regime of low sSFR. However, this would require much deeper observations, as stellar absorption kinematics are much more difficult to measure compared to H$\alpha$ emission lines.

In Figure~\ref{ssfr_mass} and Figure~\ref{ssfr_hist} we include a much deeper and more complete sample from COSMOS2015 catalog. We show that this more complete COSMOS2015 dwarf sample does cover the full dynamic range of sSFR in the FIRE sample, i.e. there could be observed galaxies with low sSFR as FIRE-2 produced. We also point out that FIRE-2 simulations do not produce high sSFR galaxies as observed in the real universe, as shown in Figure~\ref{ssfr_mass} and Figure~\ref{ssfr_hist}. Our kinematics results also show that the observed $\rm \Delta\sigma_{gas}$-sSFR relation is less tight than what FIRE-2 predicted. This tension is in the same direction as the stellar feedback in FIRE simulations, which suppresses the star-formation by disturbing the cold gas in galaxies, and may be stronger than the observations. 

In fact, \citet{Sparre:2017} used an earlier version of the FIRE simulations, FIRE-1, to compare with observed nearby galaxies in the Local Volume from \citet{Weisz:2012}, and also found that the star-forming main sequence in FIRE is systematically lower than the observations for low-mass galaxies ($\rm \log M_*/M_\odot<9.5$), especially for star-formation rates derived from H$\alpha$, which is consistent with our results shown in Figure~\ref{ssfr_mass} and Figure~\ref{ssfr_hist}. They found that the FIRE simulations over-predicted the burstiness of the star-formation histories in low-mass galaxies at $z\sim 0$, which could be due to the fact that FIRE simulations do not fully resolve the giant molecular clouds (GMCs) down to the lowest-mass GMCs. The unresolved low-mass GMCs might be able to sustain a low-level SFR. Given the large number of low-mass GMCs, the SFR in simulated low-mass galaxies might be underestimated. 

\citet{Emami:2021} found that a stronger correlation between sizes and SFR in FIRE-2 simulations than in the observations, indicating that the timescale of the bursty star-formation in FIRE is much shorter than the observations. Furthermore, several other groups \citep{Kado-Fong:2022, Jiang:2023, El-Badry:2018a, El-Badry:2018b} also showed that FIRE-2 simulations produce too many diffuse and non-disky dwarf galaxies compared to the real dwarf population in observations, also suggesting that the stellar feedback from the bursty star-formation in FIRE simulations may differ from the real universe.

We emphasize that our FIRE-2 sample is much smaller in terms of the number of individual galaxies compared to the observational samples, and the FIRE results do not include observational uncertainties. Both of these factors would push the observations to have larger scatter than in FIRE-2 simulations. Moreover, the 8 low-mass galaxies from the FIRE-2 simulations are all isolated galaxies, whereas the observational samples (both H$\alpha$ detected dwarf sample and COSMOS2015 dwarf sample) are a mixture of isolated galaxies and galaxies in groups. Since the FIRE-2 simulations are not designed to reproduce the full population of galaxies with different environments and initial conditions, we do not expect that this relatively small FIRE-2 sample would have the same sSFR distribution as the COSMOS2015 dwarf sample. The fact that the COSMOS2015 sample includes low sSFR galaxies as FIRE-2 produced shows that if we could measure the gas velocity dispersion for those low sSFR galaxies, we should be able to include them in Figure~\ref{dsigma_ssfr}. Nevertheless, our main results would not change even if we don't have those low sSFR galaxies in our H$\alpha$ detected dwarf sample, since we have the same results with all low sSFR galaxies in FIRE-2 taken out, as shown in Figure~\ref{dsigma_ssfr_cut}.

Recently, \citet{Klein:2025} analyzed $\sim$700 star-forming central galaxies in the FIREbox simulation \citep{Feldmann:2023, Benavides:2025}, which uses the same physics as FIRE-2 \citep{Hopkins:2018} but has a much larger volume (15 cMpc $h^{-1})^3$ at a lower resolution ($\sim$30 times lower than FIRE-2). \citet{Klein:2025} found that, even with this larger sample, FIREbox still fails to produce a sufficient number of low-mass disk galaxies compared to observations down to $\rm \log M_*/M_\odot>8.0$. This suggests that the discrepancy between observations and FIRE simulations for low-mass galaxies cannot be attributed solely to the limited sample size or volume.

In this work, we primarily focus on testing the relation between ionized gas kinematics and recent star-formation rates driven by the breathing mode. Both are measured within a single inflow/outflow episode, as ionized gas responds rapidly to the injection of energy and momentum from stellar winds and supernovae. However, the breathing mode scenario includes not only short-timescale processes within individual episodes, but also long-term effects driven by repeated inflow/outflow cycles \citep{El-Badry:2016}. In simulations, lasting dark matter cores are formed through many such repeated and semi-periodic inflow/outflow episodes \citep[e.g.][]{Pontzen:2012,Chan:2015,Tollet:2016,Mostow:2024}. The radial migration of stars and dark matter is influenced not only by short-timescale potential fluctuations from individual starbursts, but also by sustained bursty star-formation activities over time. Therefore, although this work supports the short-timescale gas migration predicted in the FIRE-2 simulations, it remains unclear whether the breathing mode in FIRE-2 is consistent with observations beyond individual episodes. The amplitudes and timescales of the bursty star-formation in low-mass galaxies need to be further constrained in order to better understand the feedback effects.

\section{Conclusions}\label{conclusions}
Recent hydrodynamical simulations such as FIRE-2 simulations have been using stellar feedback to produce dark matter cores as a possible baryonic solution to the ``core-cusp'' problem. The ``breathing mode'' stellar feedback in simulations produces not only dark matter cores in low-mass galaxies by fluctuating their gravitational potential, but also dynamical effects in baryons at the same time. It predicts a positive correlation between gas velocity dispersions ($\rm \sigma_{gas}$) and the specific star-formation rates (sSFR), which could be tested with low-mass galaxies in observations.

In this paper, we test the ``breathing mode'' stellar feedback in FIRE-2 simulations with a sample of 103 observed low-mass galaxies with $7.9<\rm \log M_*/M_\odot<9.6$ at $0.02<z<0.19$. We measure the $\rm \sigma_{gas}$ from H$\alpha$ with spectra obtained by Keck/DEIMOS, and derive the star-formation rates with multi-band SED fitting and H$\alpha$ emission lines.  We mock observe the simulated galaxies in the same mass range in the FIRE-2 simulations and make direct comparisons with observations. Our main results are summarized as follows. 
\begin{itemize}
\item Our observations of low-mass galaxies show a comparable range of $\rm \sigma_{gas}$ at a given stellar mass with the FIRE-2 simulations. The observed low-mass galaxies generally follow the mass-$\rm \sigma_{gas}$ relation in the FIRE-2 simulations.
\item We confirm the positive/non-zero correlation between the mass-independent gas velocity dispersions ($\rm \Delta\sigma_{gas}$) and sSFR averaged over two different timescales ($\rm sSFR_{100Myr}$ and $\rm sSFR_{10Myr}$) with observations. However, the $\rm \Delta\sigma_{gas}$-sSFR relations in observations have larger scatter than the prediction from FIRE-2 simulations. The results from $\rm \Delta\sigma_{gas}$-sSFR relations still hold for a high sSFR sub-sample in FIRE-2 simulations matched to the observed H$\alpha$ detected sample.
\item With a larger and complete photometric sample taken from COSMOS2015, we show that observed low-mass galaxies could span the full range of sSFR predicted in the FIRE-2 simulations. However, FIRE-2 simulations do not produce low-mass galaxies with high sSFR ($\rm log(sSFR_{100Myr})>-9.0\ yr^{-1}$) at low-redshifts comparing the observations.
\item Our results, based on gas kinematics and recent star-formation rates, support the existence of short-timescale dynamical effects driven by the ``breathing mode'' stellar feedback in low-mass galaxies. However, the cumulative effects of sustained, bursty star-formation over longer timescales are still not well constrained and may contribute to the discrepancies between FIRE-2 simulations and observations. Future work using detailed SED modeling with non-parametric star-formation histories for observed low-mass galaxies could provide valuable constraints on these longer-timescale effects.

\end{itemize}

Future observations using gas and stellar kinematics measured with deeper and higher spectral resolution data will reduce selection biases while enabling more stringent constraints on stellar feedback in galaxies with even lower stellar masses, as the ``breathing mode'' is predicted to be less significant for galaxies with $\rm \log M_*/M_\odot<7.0$, a regime that has not yet been observationally tested. Incorporating additional observational properties, such as galaxy sizes and stellar population gradients, will further refine our understanding of feedback-driven breathing modes and their impact on the global properties of low-mass galaxies \citep{Riggs:2024}. Larger and deeper imaging surveys using narrow or medium-band filters targeting emission lines of low-mass galaxies, such as the Merian Survey \citep[]{Danieli:2024, Luo:2024}, could provide spatially resolved ionized gas distributions \citep[]{Mintz:2024}, offering valuable insights into the processes of star-formation and stellar feedback in low-mass galaxies.

\section*{Acknowledgements}

We thank Jenny Greene and Sandra Faber for their constructive and insightful suggestions. YL thanks Weichen Wang and Adebusola Alabi for helpful discussions regarding the Keck/DEIMOS observations.

This material is based upon work supported by the National Science Foundation under Grant No. 2106839. We acknowledge use of the lux supercomputer at UC Santa Cruz, funded by NSF MRI grant AST 1828315. AW received support from NSF, via CAREER award AST-2045928 and grant AST-2107772, and HST grant GO-16273 from STScI.

Some of the data presented herein were obtained at Keck Observatory, which is a private 501(c)3 non-profit organization operated as a scientific partnership among the California Institute of Technology, the University of California, and the National Aeronautics and Space Administration. The Observatory was made possible by the generous financial support of the W. M. Keck Foundation.

We used the FIRE-2 simulations, which are publicly available \citep{Wetzel:2023} at \url{http://flathub.flatironinstitute.org/fire}.

The authors wish to recognize and acknowledge the very significant cultural role and reverence that the summit of Maunakea has always had within the Native Hawaiian community. We are most fortunate to have the opportunity to conduct observations from this mountain.

This research made use of:
  \href{http://www.scipy.org/}{\texttt{SciPy}},
      an open source scientific tool for Python (\citealt{SciPy});
  \href{http://www.numpy.org/}{\texttt{NumPy}},
      a fundamental package for scientific computing with Python (\citealt{NumPy});
  \href{http://matplotlib.org/}{\texttt{Matplotlib}},
      a 2-D plotting library for Python (\citealt{Matplotlib});
  \href{http://www.astropy.org/}{\texttt{Astropy}}, a community-developed
      core Python package for astronomy (\citealt{astropy:2013, astropy:2018, astropy:2022});
  \href{https://dynesty.readthedocs.io/en/stable/}{\texttt{dynesty}},
      a dynamic nested sampling package for estimating Bayesian posteriors and evidences (\citealt{Speagle:2020});
  \href{https://www.star.bris.ac.uk/~mbt/topcat/}{\texttt{TOPCAT}},
      a user-friendly graphical program for viewing, analysis and editing of tables (\citealt{topcat}).

\appendix
\label{Appendix}

\section{Different $\rm M_* - \sigma_{gas}$ relations}
\label{diffslopes}
In Section~\ref{sigma_gas} we have shown the the mass-$\rm\sigma_{gas}$ relations derived from FIRE-2 simulations and our H$\alpha$ detected dwarf sample at $z<0.19$. We show that our H$\alpha$ detected dwarf sample at $z<0.19$ is not complete at low-mass and low sSFR end. According to the FIRE-2 simulations, those galaxies with low sSFR also tend to have low $\rm\sigma_{gas}$ at given stellar masses due to the feedback-driven breathing mode. The incompleteness of our sample may affect the slope of the mass-$\rm \sigma_{gas}$ relation. Therefore, the results we show in Section~\ref{sigma_mass} are all based on the $\rm \Delta\sigma_{gas}$ calculating using one same mass-$\rm \sigma_{gas}$ relation derived from the FIRE-2 simulations, assuming FIRE-2 simulations produce the correct mass-$\rm \sigma_{gas}$ relation. This assumption could be tested with further observations. However, we now show that our main results do not change with the mass-$\rm\sigma_{gas}$ relation derived from our incomplete H$\alpha$ detected dwarf sample at $z<0.19$.

\begin{figure*}
\begin{center}
\includegraphics[width=15cm]{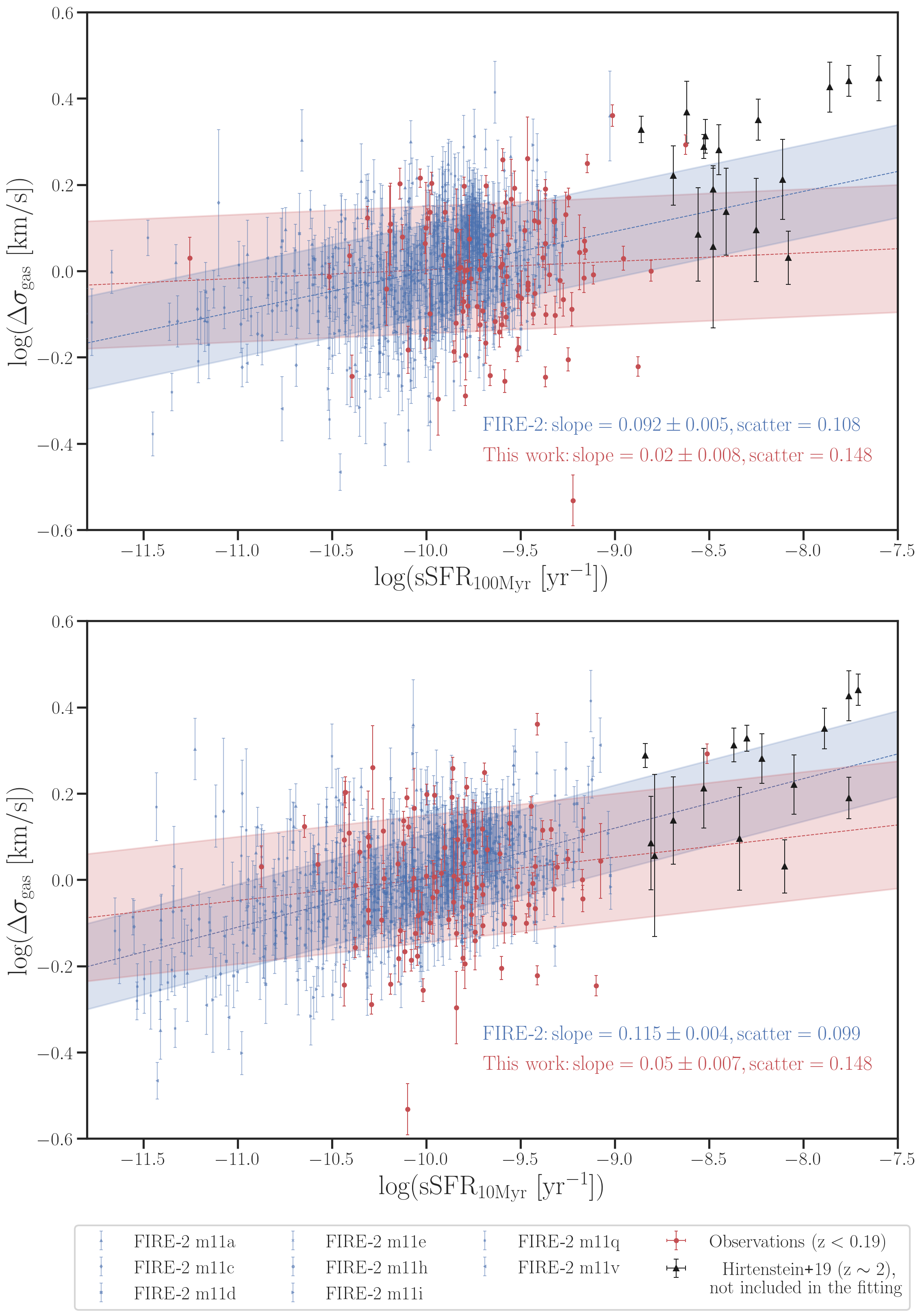}
\caption{Same as Figure~\ref{dsigma_ssfr}, but using $\rm \Delta\sigma_{gas}$ calculated with different mass-$\rm \sigma_{gas}$ relations. $\rm \Delta\sigma_{gas}$ for the observed galaxies (both the $z<0.19$ sample and \citet{Hirtenstein:2019} sample) are computed using the mass-$\rm \sigma_{gas}$ relation derived from our incomplete H$\alpha$ detected dwarf sample at $z<0.19$. With the new $\rm \Delta\sigma_{gas}$, the observed sample show a much shallower and slightly positive $\rm \Delta\sigma_{gas}$-$\rm sSFR_{100Myr}$ relation. The $\rm \Delta\sigma_{gas}$-$\rm sSFR_{10Myr}$ relation is still steeper and closer to the relation from FIRE-2 simulations, same as the results in Figure~\ref{dsigma_ssfr}.}
\label{dsigma_ssfr_diffslope}
\end{center}
\end{figure*}

\begin{figure*}
\begin{center}
\includegraphics[width=15cm]{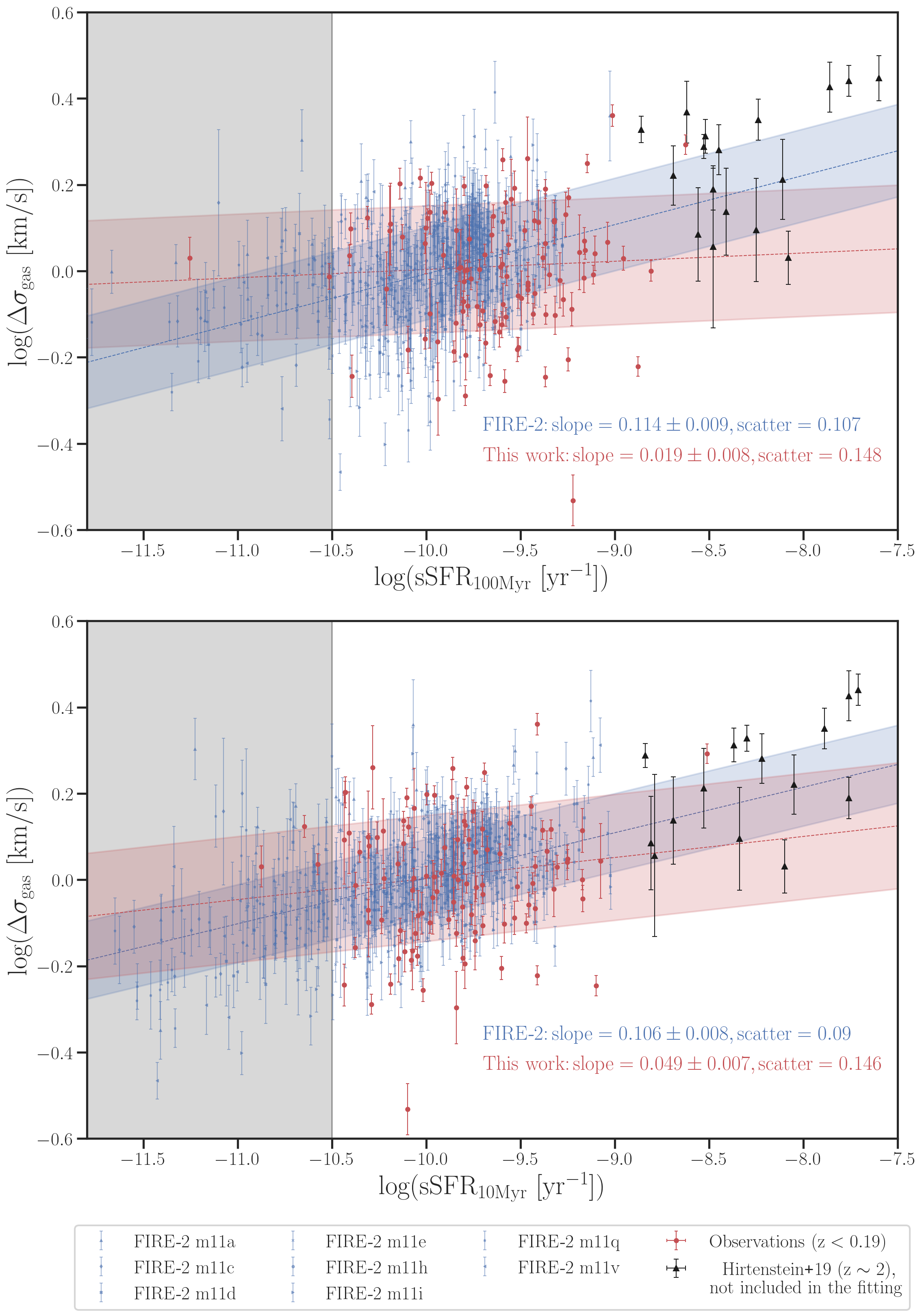}
\caption{Same as Figure~\ref{dsigma_ssfr_cut}, but using $\rm \Delta\sigma_{gas}$ calculated with different mass-$\rm \sigma_{gas}$ relations. The galaxies from observations and simulations with $\rm log(sSFR)<-10.5\ yr^{-1}$ (shaded area) are not used for fitting the relations, same as Figure~\ref{dsigma_ssfr_cut}. $\rm \Delta\sigma_{gas}$ for the observed galaxies (both the $z<0.19$ sample and \citet{Hirtenstein:2019} sample) are computed using the mass-$\rm \sigma_{gas}$ relation derived from our incomplete H$\alpha$ detected dwarf sample at $z<0.19$. With the new $\rm \Delta\sigma_{gas}$, the observed sample show a much shallower and slightly positive $\rm \Delta\sigma_{gas}$-$\rm sSFR_{100Myr}$ relation. The $\rm \Delta\sigma_{gas}$-$\rm sSFR_{10Myr}$ relation is still steeper and closer to the relation from FIRE-2 simulations, same as the results in Figure~\ref{dsigma_ssfr}.}
\label{dsigma_ssfr_cut_diffslope}
\end{center}
\end{figure*}

Figures~\ref{dsigma_ssfr_diffslope} and \ref{dsigma_ssfr_cut_diffslope} show the $\rm \Delta\sigma_{gas}$-sSFR relations as we have shown in Figure~\ref{dsigma_ssfr} and Figure~\ref{dsigma_ssfr_cut}. However, the $\rm \Delta\sigma_{gas}$ for the observed galaxies in Figure~\ref{dsigma_ssfr_diffslope} and Figure~\ref{dsigma_ssfr_cut_diffslope} are not calculated with the mass-$\rm \sigma_{gas}$ relation derived from FIRE-2 simulations. Instead, the $\rm \Delta\sigma_{gas}$ for observed galaxies are calculated using mass-$\rm \sigma_{gas}$ relation derived from our H$\alpha$ detected dwarf sample at $z<0.19$. The simulated galaxies are still using the same mass-$\rm \sigma_{gas}$ relation derived from FIRE-2 simulations. With this different $\rm \Delta\sigma_{gas}$ for the observed galaxies, the $\rm \Delta\sigma_{gas}$-$\rm sSFR_{100Myr}$ relations is much shallower, for both the full H$\alpha$ detected dwarf sample (Figure~\ref{dsigma_ssfr_diffslope}) and the sSFR limited sample (Figure~\ref{dsigma_ssfr_cut_diffslope}). The slopes are still positive, but close to 0 (flat). The $\rm \Delta\sigma_{gas}$-$\rm sSFR_{10Myr}$ relations in Figure~\ref{dsigma_ssfr_diffslope} and Figure~\ref{dsigma_ssfr_cut_diffslope} are steeper and closer to the relations from FIRE-2 simulations, same to what we have seen in Figure~\ref{dsigma_ssfr} and Figure~\ref{dsigma_ssfr_cut}.

\bibliography{breathing_mode}{}
\bibliographystyle{aasjournalv7}

\end{document}